\documentclass[10pt]{iopart}
\expandafter\let\csname equation*\endcsname\relax
\expandafter\let\csname endequation*\endcsname\relax
\usepackage{amsmath}
\usepackage{graphicx}
\usepackage{comment}
\usepackage{orcidlink}
\usepackage[switch]{lineno}

\begin{document}

\title[Self-consistent modelling and qualitative comparison of mildly relativistic RE dynamics to CFSF model]{Self-consistent modelling and qualitative comparison of mildly relativistic runaway electron dynamics with a closed flux surface formation model during tokamak startup}

\author{Y. Lee$^{1,2}$\orcidlink{0000-0003-4474-416X}, 
H.-T. Kim$^3$\orcidlink{0009-0008-2549-5624}, 
P.C. de Vries$^4$\orcidlink{0000-0001-7304-5486}, 
P. Aleynikov$^5$\orcidlink{0009-0002-3037-3679}, 
J. Lee$^6$\orcidlink{0000-0002-2353-2603}, 
K. Park$^1$\orcidlink{0009-0001-1397-2701}, 
T. Park$^1$\orcidlink{0009-0008-1471-2013}, 
J. Gwak$^1$\orcidlink{0000-0003-4767-3328}, 
G. Nam$^1$, 
W. I. Jeong$^6$, 
K.-D. Lee$^6$, 
J.-G. Bak$^6$, 
J. Jang$^6$, 
J.-W. Juhn$^6$, 
Y.-S. Lee$^6$, 
J.-K. Park$^{1}$\orcidlink{0000-0003-2419-8667} and 
Y.-S. Na$^{1*}$\orcidlink{0000-0001-7270-3846}}

\address{$^1$ Department of Nuclear Engineering, Seoul National University, Seoul, South Korea} \label{ad:1}
\address{$^2$ Nuclear Research Institute for Future Technology and Policy, Seoul National University, Seoul, South Korea}
\address{$^3$ United Kingdom Atomic Energy Authority, Culham Campus, Abingdon, Oxfordshire OX14 3DB, United Kingdom}
\address{$^4$ ITER Organization, Route de Vinon sur Verdon, CS 90 046, 13067 St Paul Lez Durance, France}
\address{$^5$ Max-Planck Institute fur Plasmaphysik, Greifswald, Germany}
\address{$^6$ Korean Institute of Fusion Energy, Daejeon, South Korea}

\ead{ysna@snu.ac.kr$^*$}
\vspace{10pt}
\begin{indented}
\item[]Mar 2026
\end{indented}

\begin{abstract}
A model for mildly relativistic Runaway Electrons (REs) is developed in a reduced-kinetic form and qualitatively compared with radiation characteristics observed in KSTAR ohmic startup. The mildly relativistic correction not only alleviates runaway current overestimation but also accounts for the partial parallel confinement of the initial runaway seed under an open-field configuration during early burn-through. The model is self-consistently integrated in the state-of-the-art predictive plasma initiation code DYON (Hyun-Tae Kim \textit{et al} 2022 \textit{Nucl. Fusion} \textbf{62} 126012), hereafter referred to as DYON-RE. DYON-RE provides an improved RE confinement model during the transition from an open to a closed magnetic configuration by employing a model-based description of closed flux surface formation validated in multi machines. We show prediction capability of DYON-RE in two representative discharges among KSTAR ohmic startups. DYON-RE reliably predicts key plasma parameters such as plasma current, density, and temperature and also implies the characteristic behavior of the radiative temperature measured by electron cyclotron emission diagnostics in agreement with experimental results. The proposed model offers a framework for designing runaway-free ohmic startup scenarios in CPD and ITER. Future experimental validation will further refine its predictive capabilities and broaden its practical application.
\end{abstract}

%
\vspace{2pc}

\submitto{\NF}
\maketitle 
\ioptwocol

\section{Introduction}
Significant generation of Runaway Electrons (REs) has long been recognized as a critical concern for localized damage to plasma-facing components during tokamak disruptions \cite{Breizman2019NF}. 
The focus on RE in tokamaks has recently been on their generation as part of the disruption process \cite{Martin-solis2017NF, Breizman2019NF}. However, it is well known that RE can also appear during the starup of the tokamak discharge \cite{Knoepfel1979NF, Vries2019NF}. 
Concerns about their detrimental effects were supported by conservative numerical predictions of runaway-driven startup failures if the particle density is too low \cite{Gribov2018EPS, Lee2024PRL}. In addition, experimental observation in the WEST tokamak indicates that REs can lead to quench of magnet \cite{Reux2021IAEA}. These findings suggest that, in reactor tokamaks, startup REs may not only hinder plasma initiation but also pose a potential threat to the device safety.

The first simulation of startup RE generation was performed by Sharma and Jayakumar \cite{Sharma1988NF} by coupling the prototype plasma burn-through model \cite{Hawryluk1976NF} with the Dreicer generation model \cite{Connor1975NF}. In their approach, the runaway velocity is inferred from the free-fall acceleration during the confinement time characterized by drift orbit loss \cite{Knoepfel1979NF}; the drift orbit loss alone has been found to be insufficient to explain modern experimental observations in middle-size tokamaks \cite{Vries2023NF, deVries2025NF}. Later studies have improved the RE models \cite{Gribov2018EPS, Hoppe2022JPP, Matsuyama2022NF, deVries2025NF}. During these model improvements, the Sharma–Jayakumar correction was simplified to the assumption that the runaway velocity equals the speed of light to estimate runaway current density. However, when electrons are formally in the runaway regime but have not yet reached relativistic speeds, this assumption can significantly overestimate the runaway current density. Meanwhile, Ref. \cite{Hoppe2022JPP} proposed an approach to estimate the RE confinement time by adapting the interpolating treatment of thermal particle confinement \cite{Kim2012NF} and inferring the parallel free-fall time following the connection magnetic field line \cite{Vries2019NF}, although it considers a single runaway population. This approach may be well suited for describing the RE transport in an open field configuration; however, in the presence of a local closed flux surface, an additional runaway population needs to be considered to capture the improved transport. In order to enhance the prediction of early runaway dynamics, we develop a \textit{mildly relativistic} runaway model\footnote{We refrain from using the term \textit{supra-thermal} to rule out supra-thermal non-runaway electrons.} and self-consistently couple it with the electromagnetic model where the magnetic configuration evolves during Closed Flux Surface Formation (CFSF).

Only recently has systematic comparison between simulation and experiment been undertaken \cite{deVries2025NF}. The first validation trial \cite{deVries2025NF} showed that, owing to non-uniqueness of solutions, free-parameter variations such as recycling coefficient and gas fueling assumption should be minimized to evaluate the reproducibility of experimental data in simulations; Ref. \cite{deVries2025NF} also showed that the simulation result is sensitive to the choice of the RE loss model. Hence, advances in the understanding and modelling of RE losses during tokamak startup would improve such simulations. This philosophy was applied to the state-of-the-art predictive plasma initiation code DYON \cite{Kim2012NF, Kim2022NF}, whose predictive capability for the operational window has been validated across multiple machines using only hardware design and control-room input data \cite{Kim2026NF}. In the validated plasma description of DYON, the mildly relativistic RE model has been self-consistently implemented (referred to as DYON-RE hereafter). To assess the model validity, we perform the qualitative comparison of DYON-RE to observations in KSTAR ohmic discharges \cite{Lee2022EPS}.

Runaway formation has different features during startup and disruption. In post thermal quench plasmas during disruption, strong collisional damping hinders the onset of kinetic instabilities due to the low plasma temperature \cite{Aleynikov2015NF}. On the other hand, during startup, although kinetic instabilities are initially suppressed by the low plasma temperature, the progressive temperature increase during early current rampup can trigger kinetic instabilities \cite{Zhang2025PoP}. This understanding provides a physical motivation of the model comparison approach in KSTAR that employs characteristics of radiation emitted by REs.

Ideally, a desirable approach to the model validation would be to demonstrate the runaway-driven burn-through failure in experiments and to reproduce the same result using DYON-RE. Currently, KSTAR is not equipped with direct measurement for runaway current such as gamma-ray camera \cite{Paz-soldan2017PRL, Paz-Soldan2018PoP} and it is therefore challenging to unambiguously attribute a burn-through failure specifically to runaway electrons. Instead, the detection of sawtooth-like pattern in Electron Cyclotron Emission (ECE) measurement can be a qualitative indicator of runaway current density \cite{Knoepfel1979NF}. 
In other words, the prediction of the onset threshold enables a rough macroscopic estimation of a lower bound on the runaway current density \cite{Aleynikov2015NF}. In addition, non-thermal effects resulting from kinetic instabilities can be exploited for comparative validation \cite{Harvey1993PoF, Liu2018PRL, Liu2018NF}. In this work, we adopt Kinetic Instability Analysis Tool (KIAT) and SYnthetic NOn-thermal electron cyclotron emission reconstruction tool (SYNO) \cite{Lee2026arXiv2} to analyze the threshold runaway density for kinetic instability onset and corresponding non-thermal effect in ECE measurement at the linear level, respectively.

In summary, the DYON-RE code will reproduce KSTAR ohmic discharges. Key model improvements are to elaborate current density models carried by mildly relativistic REs and to capture the transition between RE losses in open and closed field line structures. This paper is organized as follows. Section \ref{sec:BI_sRE} provides a brief introduction to startup REs for readers with expertise in either startup physics or RE physics, but not necessarily both. Readers already familiar with startup REs may skip this section. In Section \ref{sec:MR_model}, we develop a mildly relativistic runaway electron model in a reduced-kinetic form. In Section \ref{sec:coupling}, coupling of the runaway model with DYON is described after briefly revisiting the plasma burn-through model. In Section \ref{sec:validation}, we assess the model validity by qualitatively comparing DYON-RE in KSTAR.

\section{Brief introduction of startup runaway electrons \label{sec:BI_sRE}}
In Section \ref{ssec:RSRE}, we briefly introduce the definition of the fluid REs consistent with time-dependent kinetic simulations as the background of a model development of startup REs with the mildly relativistic correction. A comprehensive review of general runaway electron physics can be found in Ref. \cite{Breizman2019NF}, while a review focusing on the startup context is given in Ref. \cite{Vries2023NF} and Sec. 2.4.3. of Ref. \cite{Na2025NF}.

Previous studies have coupled RE models to plasma startup models using the macroscopic (zero-dimensional) description of plasma parameters \cite{Sharma1988NF, Gribov2018EPS, Hoppe2022JPP,  Matsuyama2022NF, deVries2025NF}. However, several aspects have not yet been fully explored, including where the description can be usefully adopted, how the evolution of plasma parameters is modelled and how the simplifying treatments affect the runaway electron modelling. Therefore, we also revisit basic principles and issues that are inherently embedded in the macroscopic (zero-dimensional) description in Section \ref{ssec:RPBT}. The purpose is not to suggest that all of these issues will be resolved within the present work, but rather to clarify which aspects can be meaningfully addressed here and which lie beyond the scope of this study.

\subsection{Definition of runaway electrons in the RE fluid model \label{ssec:RSRE}}
Fluid description of disruption runaway electrons often assumes RE velocity to be the speed of light because of the effect of the strong electric field induced by the rapid current quench. However, from the perspective of electron kinetics, the definition of runaway electrons needs to be more specific than simply particles moving close to the speed of light. If runaway electrons are simply defined as electrons whose momentum exceeds the critical momentum $p_c$, at which the electric acceleration balances collisional friction, then a significant fraction of the runaway population can remain classical when $p_c$ lies well below the relativistic regime.

The RE fluid models are typically obtained by solving a time-dependent electron kinetics with the steady-state assumption on background plasmas \cite{Gurevich1961JETP, Connor1975NF, Rosenbluth1997NF}. As a result, one may overlook what definition of runaway electrons is implicitly presumed in the models. For example, in the case of Dreicer generation \cite{Connor1975NF} (primary generation), which is known to play a leading role in RE formation during startup \cite{Na2025NF}, the generation rate is calculated from the steady-state particle flux into the runaway regime in phase space under an applied electric field. Consequently, the resulting runaway electron generation rate becomes independent of the specific choice of the boundary in momentum space used to define REs, since the phase space flow is effectively incompressible due to the particle conservation under small angle Coulomb scatterings.

\begin{figure}
    \centering
    \includegraphics[width=0.5\textwidth]{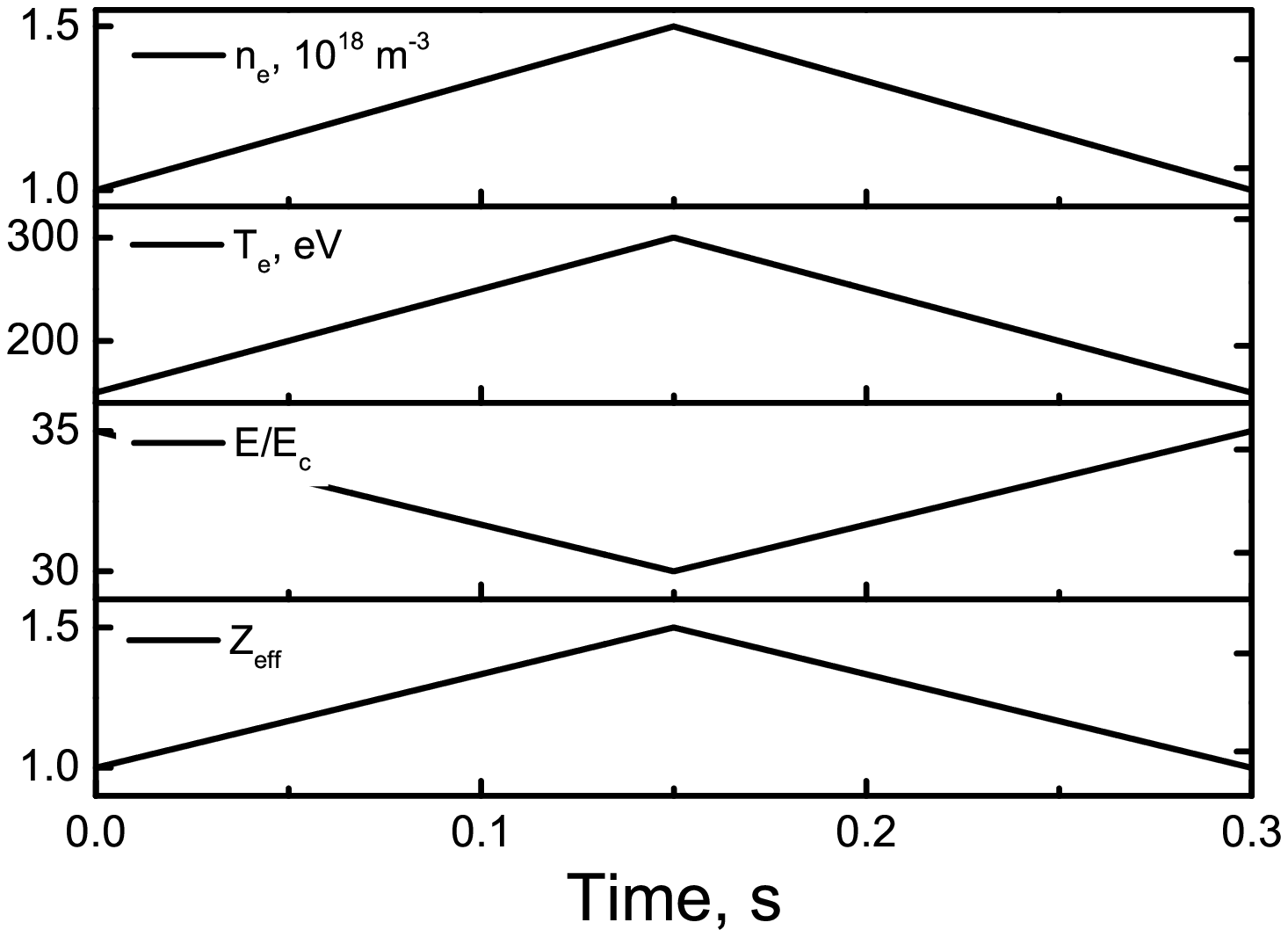}
    \caption{Time evolutions of plasma parameters relevant to KSTAR startup (early rampup). Adapted from Ref. \cite{Lee2023NF}. © IAEA/IOP Publishing. Used with no objection from the publisher.}
    \label{fig:param_evol}
\end{figure}

\begin{figure}
    \centering
    \includegraphics[width=0.5\textwidth]{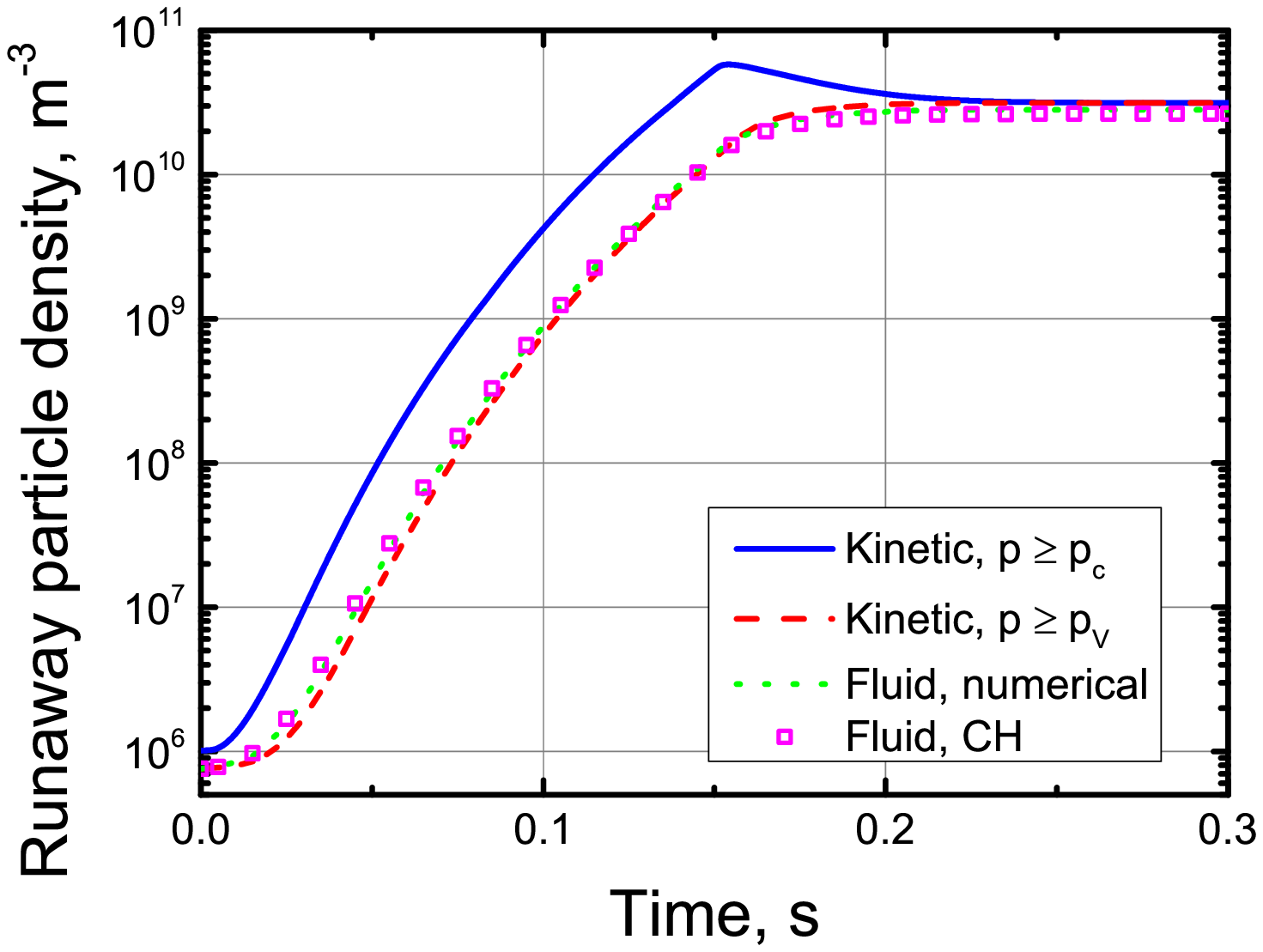}
    \caption{Time evolutions of the RE density from kinetic simulations under plasma parameter variations in Fig. \ref{fig:param_evol} where $n_{RE}$ is evaluated above $p_{c}$ (blue solid line) and $p_{V}$ (red dashed line). Green dotted line and magenta square marker are the fluid model predicted by kinetic simulation and the Connor-Hastie formula \cite{Connor1975NF}, respectively. Adapted from Ref. \cite{Lee2023NF}. © IAEA/IOP Publishing. Used with no objection from the publisher.}
    \label{fig:nRE_full}
\end{figure}

However, a proper choice of the runaway boundary is required to infer the RE generation rate when plasma parameters change in time \cite{Lee2023NF}. Figure \ref{fig:param_evol} shows the time variation of the plasma parameters in the example problem set up in Ref. \cite{Lee2023NF}, which is relevant to the KSTAR startup plasma. Here, $E_c \equiv \frac{e^3 n_e \ln \Lambda_{free}}{4\pi \varepsilon_0^2 mc^2}$ is the critical electric field and it has the following relationship with $p_c$ by $(\frac{E}{E_c})^2 = 1 + \frac{1}{p_c^2}$. The test problem is constructed by parametrically ramping the plasma parameters up through an early-rampup phase, where the primary generation grows exponentially, and then deliberately ramping them down so that the primary generation weakens again. This design ensures that any definition may be used after the RE beam has formed, whereas during its formation the correct definition is required to demonstrate consistency between the fluid and kinetic models.

Figure \ref{fig:nRE_full} demonstrates that if the runaway boundary is defined as $p_c$ (blue curve) using the time evolving parameters in Fig. \ref{fig:param_evol}, the RE density calculated from kinetic simulation is higher by several factors than the values from the analytic (CH) and numerical fluid models, although the final saturated values are consistent.

\begin{figure}
    \centering
    \includegraphics[width=0.5\textwidth]{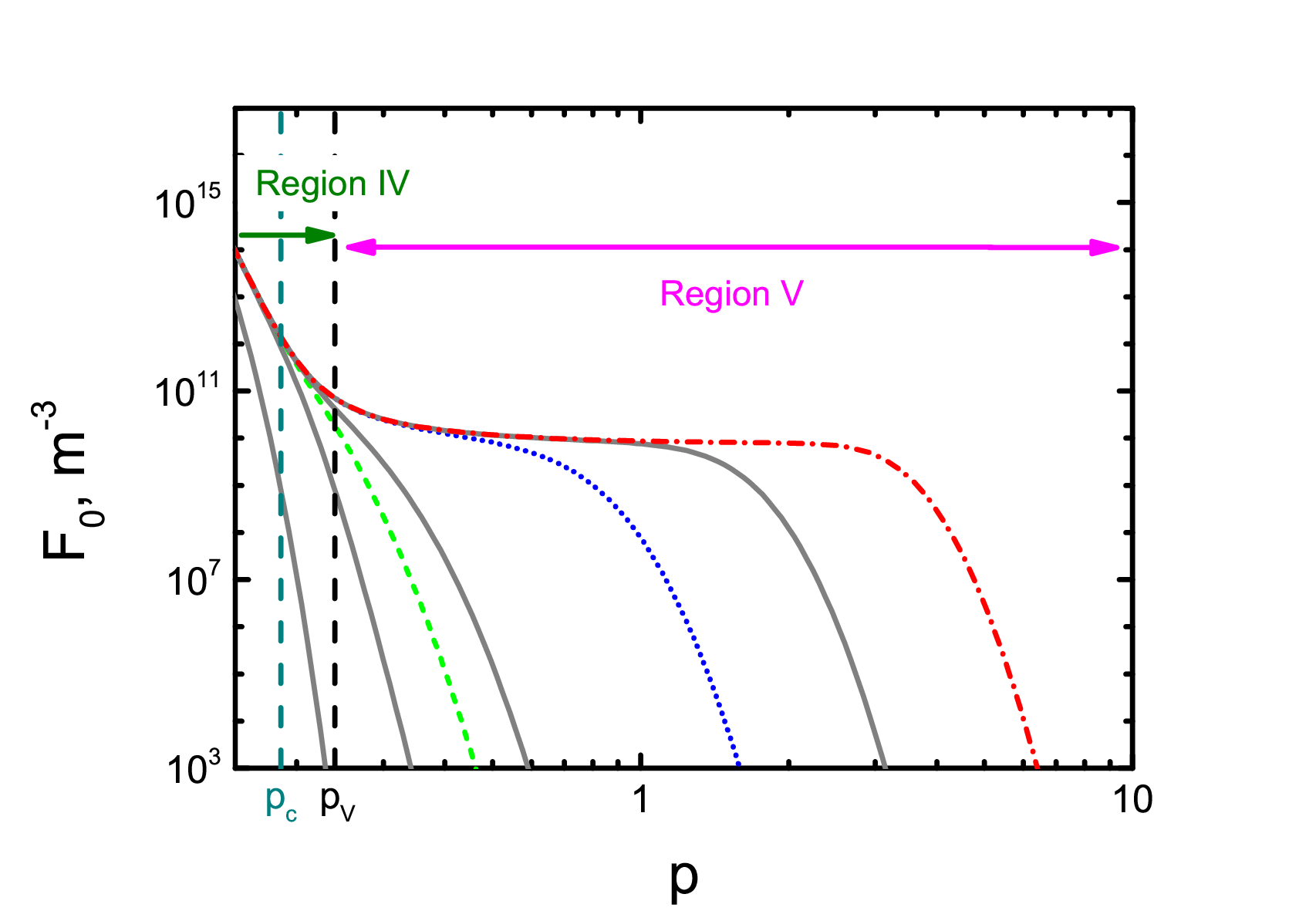}
    \caption{Snapshots of one-dimensional distributions $F_0$ such that $\int F_0 dp = n_e$ from kinetic simulations at selected time points. Plasma parameters are $n_e = 10^{18}\, m^{-3}$, $E/E_c=30$, $T_e=300 \, eV$ and $Z_{eff}=2$. Green, blue and red curves correspond to $1\times$, $2.5\times$ and $20\times$ collisional times measured at $p_c$. Grey curves denote time between them. Adapted from Ref. \cite{Lee2023NF}. © IAEA/IOP Publishing. Used with no objection from the publisher.}
    \label{fig:pV}
\end{figure}

Under the classification of phase-space regions considered in the Connor-Hastie work \cite{Connor1975NF}, a thin, singular layer forms in the vicinity of $p_c$ (indicated by Region IV in Fig. \ref{fig:pV}). Within this layer, particle motion is not governed by electric acceleration but by pitch-angle scattering and energy diffusion (see Appendix C.2 in Ref. \cite{Lee2025phD}). The "free-acceleration" condition that characterizes runaway electrons is satisfied in the region where the particle momentum is sufficiently greater than $p_c$. For instance, the one-dimensional momentum distribution $F^0$ in Fig. \ref{fig:pV} exhibits a plateau-like structure (indicated by Region V), which results from the continuous acceleration of runaway particles. These two regions have an overlapping domain in which the upward particle flux from the thin layer (IV) to the runaway region (V) can be matched. We refer to a matching boundary within this domain as the region-based critical boundary, $p_V$. After redefining the runaway boundary as $p_V$, the time evolution of the RE density predicted by the kinetic simulation (red curve in Fig. \ref{fig:nRE_full}) becomes consistent with the fluid simulation results. This example has an important implication for the development of a mildly relativistic model: the fluid model of primary RE generation describes the rate at which REs are born at $p_V$, rather than at $p_c$.

\subsection{Revisit basic principles and existing issues in global plasma burn-through model \label{ssec:RPBT}}
\subsubsection{Reconciliation with first principle simulation \label{sssec:3.1.1}}
Although breakdown and burn-through differ in their original meanings, the expansion of their coverage has led to an overlap between the post breakdown \cite{Yoo2018Nat} and early burn-through phases \cite{Kim2022NF}. The issue is that, because the origins of these conceptual extensions are distinct, the same plasma has been described using different terminologies and modeled within different frameworks. This constitutes a domain that clearly requires mediation and reconciliation.

According to Yoo et al. \cite{Yoo2018Nat}, using the two-dimensional particle-in-cell BREAK code \cite{Yoo2017CPC}, the breakdown phase consists of the Townsend avalanche phase followed by the so-called E$\times$B turbulent mixing phase. When the electron density reaches a critical level (around $O(10^{12})\, \mathrm{m^{-3}}$ in KSTAR), a self-generated electric field enhances perpendicular transport to the background magnetic field, and the resulting inhomogeneity in the perpendicular plasma profile delays electron density growth. Here, the post-breakdown phase refers to the later stage of the breakdown phase, when the electron density reaches a level of $10^{15\text{--}16},\mathrm{m^{-3}}$.

Although the plasma burn-through model often assume plasma parameters are uniform, this does not necessarily imply that the model is incompatible with nonuniform profile; rather, it may still be usefully applied over a broader range of conditions. For instance, in uniform plasmas, the ionization rate can be written as a function of $<n_e>$, $<n_D>$, $<T_e>$
\begin{equation}
    <n_e n_D \mathcal{R}_{iz}(T_e)> = <n_e> <n_D> \mathcal{R}_{iz}(<T_e>)
\end{equation}
where $<\cdots>$ is a volume averaging operator and $\mathcal{R}_{iz}\equiv<\sigma v>_{iz}$ is the ensemble averaged cross section for ionizing collisions. In non-uniform plasmas, a direct equality is not satisfied 
\begin{equation}
    <n_e n_D \mathcal{R}_{iz}(T_e)> \neq <n_e> <n_D> \mathcal{R}_{iz}(<T_e>)
\end{equation}
but an approximation still would be possible
\begin{equation}
    <n_e n_D \mathcal{R}_{iz}(T_e)> \approx <n_e> <n_D> \mathcal{R}_{iz}(\frac{<T_e>}{c_T})
\end{equation}
by introducing a factor $c_T$ that measures a profile inhomogeneity effect in atomic reactor coefficients. 

On the other hands, the evolution of electromagnetic field energy has been partially considered by the current circuit model in the plasma burn-through model \cite{Ejima1982NF}. The complete application of the so-called RLC circuit is 
\begin{equation}
    V_{loop} I_p = R_p I_p^2 + \frac{1}{2}\frac{d}{dt}(L_p I_p^2) + \frac{1}{2}\frac{d}{dt}(C_pQ_p^2). \label{eq:RLC}
\end{equation}
where $V_{loop}$ is the external loop voltage, $I_p$ is the plasma current, $R_p$ is the plasma resistivity, $L_p$ is the plasma inductance, $C_p$ is the plasma capacitor and $Q_p$ is the plasma charge. When Coulomb collisions dominate collisional dynamics, the time derivative of electric field energy charged in the plasma capacitor is negligible. Yet, as demonstrated by Yoo et al. \cite{Yoo2018Nat}, a small fraction of charge becomes gradually separated as a result of the Debye shielding process and the term $\frac{1}{2}\frac{d}{dt}(C_p Q_p^2)$ is not negligible during the post breakdown. In other words, a portion of the charge generates a self-generated (or self-consistent) electric field to compensate for this separation to satisfy quasi-neutrality. 
While the behaviors in these two asymptotic regimes are well understood, it remains unclear when the term $\frac{1}{2}\frac{d}{dt}(C_p Q_p^2)$ becomes negligible. Therefore, a dedicated study is required to quantify the validity of the burn-through model assumption that neglects the evolution of electric field energy in the energy conservation relation and its effect on the background electron temperature and resistive electric field that drive the RE primary generation, 
which is clearly beyond a scope of this work.

\subsubsection{Sensitivity of modelled runaway dynamics to prefill gas pressure \label{sssec:p0}}
The plasma burn-through model has reported a successful demonstration of prediction capability in operating space by introducing a proportional factor of prefill gas pressure $c_{p_0}$ that maps the measurement (in a diagnostic location) to the simulation input (in a center of vacuum vessel) only once, not manipulated several times. The rationales for choice of $c_{p_0}$ relied on an estimate, since direct measurement at the center of the vacuum vessel is difficult. The estimate would be reasonable if the density evolutions were consistent between the burn-through model and measurement \cite{Kim2022NF, Lee2023PS, Kim2024NF, Chen2025NF, deVries2025NF}. According to this, all present-day tokamaks need the proper $c_{p_0}$ to be less than 1, including MAST\cite{Kim2022NF}, MAST-U\cite{Kim2024NF}, VEST\cite{Yun2023IAEA,Kim2026NF}, DIII-D\cite{Kim2026NF}, KSTAR\cite{Lee2023PS} and JET\cite{deVries2025NF}; only EAST\cite{Chen2025NF} exceptionally reported $c_{p_0}$ higher than 1.

However, other research findings in KSTAR \cite{Yoo2018Nat} and VEST \cite{Jeong2023FED} have raised questions regarding this conclusion. According to BREAK, $c_{p_0}$ should be higher than 1 to observe the successful post-breakdown in KSTAR \cite{Yoo2018Nat}. This is consistent with the Mol-flow+ simulation result \cite{Kersevan2009JVS} analyzed in the VEST tokamak \cite{Jeong2023FED}, suggesting that the natural molecular flow form from a high-pressure to low-pressure region and thereby the actual prefill gas pressure at the vacuum center be higher than that measured in front of a gas pump.

Recall Sec. \ref{sssec:3.1.1} that the plasma burn-through model has assumed uniform plasma parameter ($c_T=1$) in evaluating atomic reaction rate, which means the model has described an accelerated ionizing avalanche process and a reinforced radiative barrier to overcome. Although introducing $c_T> 1$ somewhat delayed a density growth and increased a proper value of $c_{p_0}$, the chosen $c_{p_0}$ would be still far less than 1. Therefore, we explicitly disclose that the prefill gas pressure in our KSTAR simulation is likely still lower than the actual value. This does not necessarily imply that the same holds for devices other than KSTAR and VEST.

This issue would not be critical for operating space prediction but crucial for prediction of startup RE generation. If the plasma burn-through model underestimated a value of prefill gas pressure and missed prolongation of the breakdown delay, (1) the lower neutral density would facilitate the stronger runaway electron generation across overall startup, (2) the rapid increase in the ionization fraction would hinder the runaway electron generation due to strong Coulomb collisions during burn-through (See Fig. 3 in Ref. \cite{Lee2024PRL} showing an effect of the ionization fraction on the primary generation rate) and (3) the increased temperature and low electron density after successful burn-through with the low prefill pressure would drive the strong runaway electron generation. 
That is, an accurate error quantification of the RE generation rate is difficult during the burn-through
because the effects of (1) and (2) somewhat cancel out each other, whereas it is modeled as if RE formation were relatively easy during the early rampup. From this perspective, the introduction of the coefficient $c_T$ in this study should be understood as an attempt to deal with the issues, although not fully resolving, by moderately increasing the allowable prefill pressure, slightly delaying the ionizing avalanche, and raising the early rampup density in the simulation.

From an experimental perspective, Ref. \cite{Vries2023NF} reported that the dominant factor governing the generation of startup REs is not the prefill pressure, but the evolution of particle density. For instance, in WEST and JET, startup REs were observed at excessively high prefill pressures, whereas in other devices they occurred at pressures that were too low. In startup RE modelling, the evolution of electron density is influenced by the prefill gas pressure, as well as the waveforms of the recycling coefficient and external gas puffing. For the reproduction of experimental results as done in Ref. \cite{deVries2025NF} and the present work, the electron density level can be predicted with a proper choice of these waveforms compensating although the prefill gas pressure is set lower than the experimental value. However, in predictive simulations, it is difficult to determine a priori whether such compensation is necessary, and if so, how it should be implemented; a level of electron density was reliably reproduced with reasonable choices of recycling coefficients in MAST \cite{Kim2022NF}, whereas it is difficult to achieve a consistent explanation across multiple datasets in KSTAR, although it appears feasible for a few selected discharges \cite{Lee2023PS}. Therefore, in connection with experimental observations, an accurate specification of the prefill gas pressure is ultimately required to reliably reproduce the electron density evolution for the predictive purpose.

\subsubsection{Inherent inhomogeneity in an electric field during startup \label{ssec:inhomogeneity}}
This subsection aims to clarify three points. First, it conveys the startup RE context to those familiar with disruption RE modelling by clarifying how a one-dimensional self-consistent electric field obtained by solving the current diffusion equation reduces to a zero-dimensional self-consistent electric field in the circuit equation. Second, we show that a local electric field is inhomogeneous when the current rampup rate is finite. Third, we quantify an error arising from the use of the circuit equation to identify when the zero-dimensional simplification is reasonable, and when it is not. To clearly present these three points, we consider a simple and idealized situation: a circular plasma in a tokamak geometry $(r, \theta, \phi)$ with a large aspect ratio limit $a/R_0 \ll 1$, assuming a uniform current density profile. This simplification is sufficient to show inhomogeneity in an electric field profile without the full transport model as considered in Ref. \cite{Esposito1996PPCF}.


In a one-dimensional runaway current modelling, the self-consistency in electric field is accounted for by the current diffusion equation that originates from Maxwell equation. In this case, the resulting toroidal electric field is a function of radial position, 
\begin{equation}
    E_\phi = E_{loop} - E_{ind} + \frac{\mu_0 }{4}\frac{d{j}_\phi}{dt} (r^2-a^2) \label{eq:Ephi}
\end{equation}
where $E_{loop}\equiv V_{loop}/2\pi R_0$ with the vacuum loop voltage $V_{loop}$, $E_{ind}\equiv L_e\frac{d{I}_p}{dt}$ with the plasma current $I_p$, $L_e \equiv L_p - L_i$ and $L_i \equiv \mu_0 R_0 \frac{l_i}{2}$; $L_e$ and $L_i$ do not change in time for this simplified problem. Note that $E_\phi$ is a one-dimensional self-consistent electric field since it approximately satisfies the current diffusion equation, i.e. $\vec{\nabla}^2 E_\phi = \mu_0 \frac{dj_\phi}{dt}$,
\begin{equation}
    \vec{\nabla}^2 E_\phi = \frac{1}{r(R_0+r)} \frac{\partial}{\partial r} r(R_0+r) \frac{\partial}{\partial r} E_\phi \approx \mu_0 \frac{dj_\phi}{dt} + O(\frac{r}{R_0}).
\end{equation}

However, a zero-dimensional modelling often considers the self-consistency by taking a resistive electric field $E_{res}\equiv I_p R_p / 2\pi R_0$ from the RL circuit equation,
\begin{equation}
    V_{loop} I_p = V_{res} I_p + \frac{1}{2}\frac{d}{dt}(L_p I_p^2), \label{eq:RL}
\end{equation}
where we assume that runaway current is negligible for simplicity. According to the Poynting theorem \cite{Ejima1982NF, Lloyd1991NF}, the resistive voltage $V_{res}$ is defined as
\begin{equation}
    V_{res} \equiv \frac{1}{I_p} \int_{V_p} \vec{j} \cdot \vec{E} dV
\end{equation}
where $\int_{V_p}$ denotes integral over the plasma volume $V_p$ and $\vec{E}$ is a local electric field vector ($E_\phi = \vec{E}\cdot \hat{\phi}$). By definition, a resistive electric field $E_{res} \equiv V_{res}/2\pi R_0$ is an averaged value of a self-consistent electric field $E_\phi$, i.e.
\begin{equation}
    E_{res} = E_{avg} \equiv \frac{1}{\pi a^2} \int_0^a 2\pi r E_\phi dr \label{eq:E_rad_avg}
\end{equation}
when $j_\phi$ is uniform. Thus, the zero-dimensional $E_{res}$ from startup circuit modelling is the radial average of the one-dimensional $E_\phi$ from disruption current diffusion modelling, clarifying how the latter reduces to the former while retaining self-consistency in an averaged sense.

\begin{figure}[ht]
\centering
\includegraphics[width=0.5\textwidth]{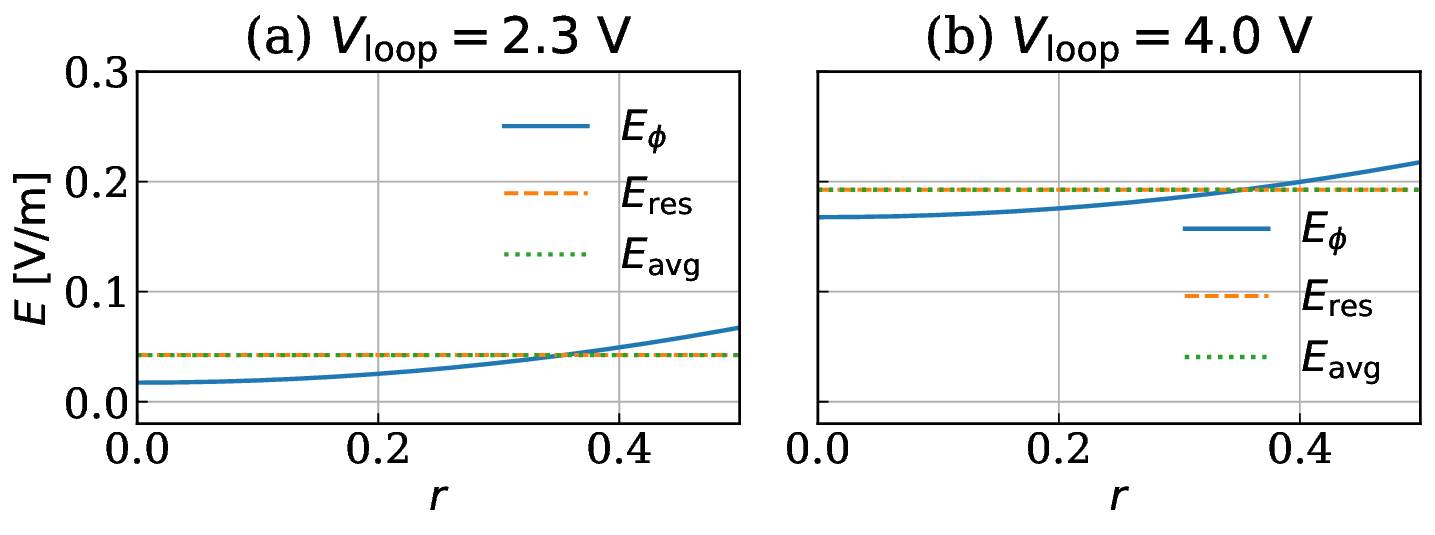}
\caption{Comparison of electric fields, $E_\phi$, $E_{res}$ and $E_{avg}$ with $V_{loop}=2.3 \ V$ (a) and $4.0 \ V$ (b). $V_{loop}$ in (a) is closer to $V_{ind} \approx 1.82 \, V$.}
\label{fig:Er}
\end{figure}

Figure \ref{fig:Er} shows radial profiles of $E_\phi$, $E_{res}$ and $E_{avg}$ under the KSTAR-relevant condition where $dI_p / dt = 0.5 \ MA\cdot s^{-1}$, $R_0 = 1.8 \ m$ and $a = 0.5 \ m$. This confirms that the radial average of $E_\phi$ coincides with $E_{res}$ regardless of $V_{loop}$, where $E_{res}$ is obtained from the circuit equation Eq. \ref{eq:RL} and $E_{avg}$ from averaging Eq. \ref{eq:Ephi}  as in Eq. \ref{eq:E_rad_avg}. In this specific example, $V_{ind} \approx 1.82 \ V$. When $V_{loop}$ far exceeds $V_{ind}$ (relevant for the burn-through phase), the relative difference is small, justifying the uniform-$E_\phi$ approximation. However, as $V_{loop}$ approaches $V_{ind}$ (relevant for the early current rampup phase), i.e. $O(V_{res}) \approx O(L_i \frac{dI_p}{dt})$, the relative difference between $E_\phi$ and $E_{res}$ becomes significant as shown in Fig. \ref{fig:Er}(a), and $E_{res}$ is no longer a suitable parameter for RE acceleration. We conclude that use of $E_{res}$ limits the range of validity of the startup runaway model to a very narrow one, e.g. up to the burn-through phase, unless the profile effect is considered well. To extend applicability of the circuit model to the early rampup, we introduce a multiplication factor $c_{n\Lambda}$ that replaces $n_e \ln \Lambda_{free}$ with $c_\Lambda n_e \ln \Lambda_{free}$ in the critical and Dreicer electric field shown later (Eqns. \ref{eq:Ec} and \ref{eq:Ed}).

\begin{figure}[ht]
\centering
\includegraphics[width=0.5\textwidth]{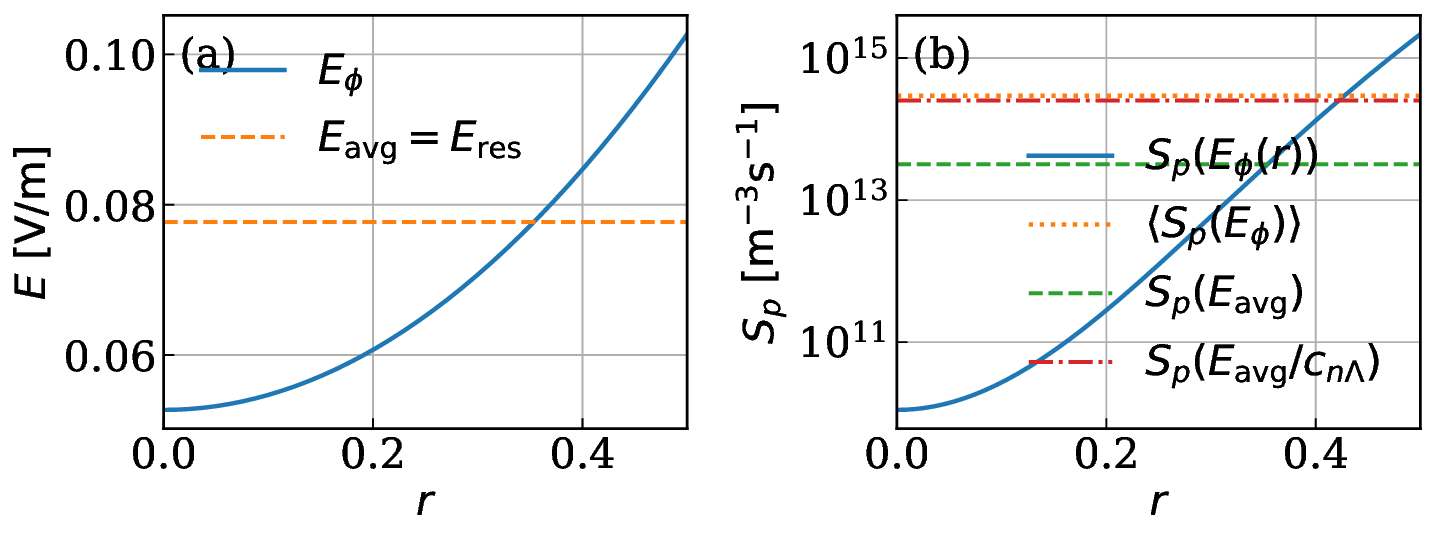}
\caption{Effect of electric field inhomogeneity on the Dreicer generation rate $S_p$. (a) shows an electric field and (b) shows $S_p$. $V_{loop}=2.7 \, \mathrm{V}$, $T_e = 250 \, \mathrm{eV}$, $n_e=2\times10^{18} \, \mathrm{m}^{-3}$ and $c_{n\Lambda} = 0.88$.}
\label{fig:prof_Sp}
\end{figure}

Note that the primary generation model of runaway electrons is also sensitive to other plasma parameters \cite{Gurevich1961JETP} and profiles \cite{Eriksson2004PRL}. Such a sensitivity was discussed in Refs. \cite{Sharma1988NF, Matsuyama2022NF, Hoppe2022JPP, Lee2023NF, deVries2025NF} in the context of startup runaway modelling. Let $S_p$ be the Dreicer generation rate \cite{Connor1975NF}. Figure \ref{fig:prof_Sp} demonstrates a significant sensitivity of $S_p$ to an electric field profile. This clarifies that, even though we assume the uniform profiles of $n_e$, $T_e$ and $j_\phi$, an inherent inhomogeneity in $E_\phi$ can produce a stiff variation in $S_p$ across the radial coordinate (blue curve in (b)). Accordingly, the direct evaluation of the averaged field $S_p(E_{avg})$ (green curve)  underestimates the volume-averaged $\langle S_p(E_\phi)\rangle$ (orange curve) significantly. Our introduction of $c_{n\Lambda}$ (red curve) alleviates the error sufficiently in this specific case, but does not necessarily do so across the broad parametric range.

\section{Reduced kinetic model of mildly relativistic runaway electrons \label{sec:MR_model}}
For the model development of mildly relativistic REs, four main findings strengthen physical foundation that we build up on the established RE physical basis \cite{Breizman2019NF}. (1) The fluid description of RE density is appropriate during a tokamak startup \cite{Lee2023NF}. (2) The Dreicer generation model predicts the particle flux across the region-based runaway boundary $p_V$, not the force-free critical momentum $p_c$ in a dynamic scenario \cite{Lee2023NF}. (3) The non-diffusive Dreicer generation should be accounted for during the early startup, particularly when the critical energy is close to the characteristic ionization energy \cite{Lee2024PRL,Lee2026JPP}. (4) Multi-fluid RE model that initializes a seed RE momentum as $p_V$ and evolves their momentum using the test particle method at every time step shows a substantial agreement with the kinetic description. We will demonstrate (4) in Sec. \ref{ssec:MFM} and show that reducing it into a single fluid form significantly improves the numerical efficiency while still preserving accuracy comparable to the multi-fluid free-fall calculation in Sec. \ref{ssec:SFM}.

The RE model presented in this section will be coupled with the DYON code in Sec. \ref{sec:coupling}. Here, the term electric field is used in a general sense and is not necessarily associated with the DYON-RE model.

\subsection{Multi fluid model of mildly relativistic runaway current \label{ssec:MFM}}
Ideally, the RE current can be precisely modelled by the multi-fluid RE populations. That is, one can consider the velocity evolution of each fluid REs that were born at each time step. Then, the RE current density $j_{RE}$ at the $i+1$-th time step is the sum of the current density carried by the fluid REs born at the $i'$-th time step ($i'\leq i$) with their velocity measured at the  $i+1$-th time step. The resulting form is
\begin{equation}
    j_{RE,i+1} = ec\sum_{i' \leq i} \Delta n_{RE,i'+\frac{1}{2}} \beta_{i'+\frac{1}{2}}(t_{i+1}), \label{eq:j_mf}
\end{equation}
where $\Delta n_{RE,i'+\frac{1}{2}}(t_{i+1})$ measures density of runaway electrons, born at the $i'$-th time step, at the $i+1$-th time step and $\beta_{i'+\frac{1}{2}}(t_{i+1})$ is their normalized mean velocity ($\beta \equiv v/c$). In our convention, the integer index $i$ means physical quantity associated with the $i$-th time step whereas the half-integer index $i+\frac{1}{2}$ indicates a physical quantity defined between time steps $i$ and $i+1$. Evolutions of $\Delta n_{RE,i'+\frac{1}{2}}(t_{i+1})$ can be tracked by considering formation and loss of REs,
\begin{equation}
\begin{split}
    \Delta &n_{RE, i'+\frac{1}{2}} (t_{i+1}) \\
    &=
    \begin{cases}
        \Big[ S_{p, i+\frac{1}{2}} + n_{RE,i}(t_{i}) \gamma_{ava,i+\frac{1}{2}} \Big] (t_{i+1}-t_i) \text{ if } i=i'\\
        \Delta n_{RE,i'+\frac{1}{2}} (t_i) \Big( 1 - \frac{1}{\tau_{RE,i+\frac{1}{2}}} (t_{i+1}-t_i) \Big) \text{ if } i>i'
    \end{cases}. \label{eq:dn_mf}
\end{split}
\end{equation}
where $S_p$ is the primary generation rate, $\gamma_{ava}$ is the avalanche growth rate and $\tau_{RE}$ is the confinement of REs. 

By construction, the total RE density evolves consistently with the general macroscopic evolution of runaway density \cite{Martin-solis2017NF}:
\begin{equation}
\begin{split}
    n_{RE,i+1}&(t_{i+1}) = \sum_{i' \leq i} \Delta n_{RE,i'+\frac{1}{2}}(t_{i+1}) \\
    &= n_{RE,i} (t_i) + (t_{i+1}-t_i) \\
    &\times \Big[ S_{p, i+\frac{1}{2}} + n_{RE,i}(t_{i}) \Big( \gamma_{ava,i+\frac{1}{2}} - \frac{1}{\tau_{RE,i+\frac{1}{2}}} \Big) \Big].
\end{split}
\end{equation}
Meanwhile each population has RE velocity with mildly relativistic correction. To estimate this, we initialize the RE velocity using $p_V$ when REs are born ($i=i'$) and describe the RE acceleration $(\frac{d\beta}{dt})_{i'+\frac{1}{2}}$ later ($i>i'$),
\begin{equation}
    \beta_{i'+\frac{1}{2}}(t_{i+1}) =
    \begin{cases}
        \beta_V \equiv \frac{p_V}{\sqrt{p_V^2+1}}  &  \text{if } i=i'\\
        \beta_{i'+\frac{1}{2}}(t_{i}) + (\frac{d\beta}{dt})_{i'+\frac{1}{2}} (t_{i+1}-t_{i})  &  \text{if } i>i'
    \end{cases}. \label{eq:beta_mf}
\end{equation}
In the test particle method, we only consider electric force and collisional friction force associated with the leading order slowing-down in electron kinetics to infer $(\frac{d\beta}{dt})_{i'+\frac{1}{2}}$ 
\begin{equation}
    \frac{d\beta}{dt} = \frac{e(E-\frac{E_c}{\beta^2})}{ m\gamma^3} \label{eq:TPM}
\end{equation}
where $E$ is the electric field, $E_c \equiv \frac{e^3 n_e \ln \Lambda_{free}}{4\pi \varepsilon_0^2 mc^2}$ is the effective critical field and $\frac{dp}{dt}=\frac{d}{dt}\frac{\beta}{\sqrt{1-\beta^2}} = \gamma^3 \frac{d\beta}{dt}$ is used. This equation is equivalent to the equation of motion of a relativistic fluid and simplifies the parallel momentum evolution $\beta_\| = \beta$, which is consistent with the definition of $p_c$ when $\frac{d\beta}{dt} = 0$. Also, radiative drag is neglected since its effect is less significant for mildly relativistic electrons than collisions.

\begin{figure}[ht]
\centering
\includegraphics[width=0.5\textwidth]{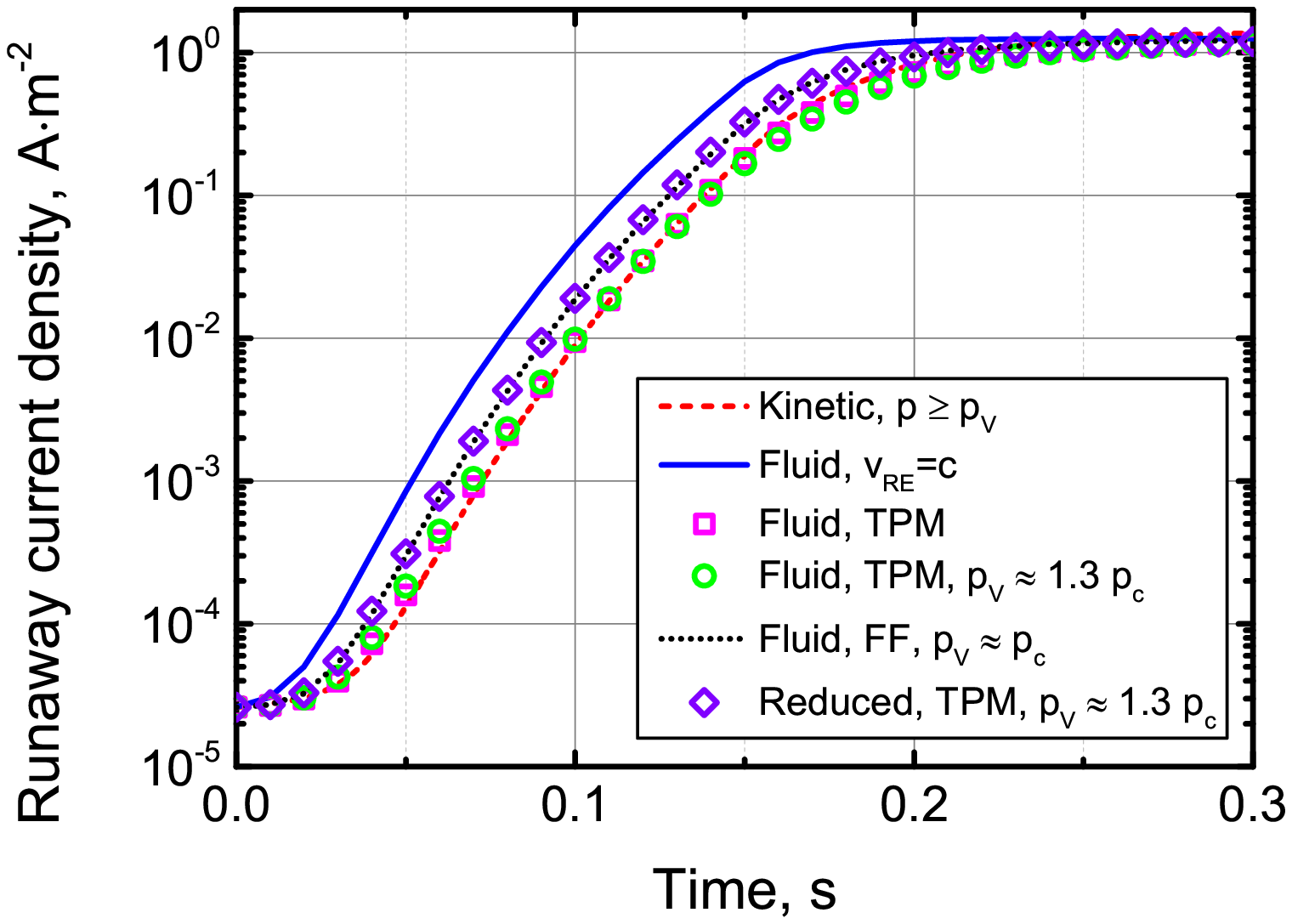}
\caption{Runaway current density estimated from the kinetic simulation (red) and fluid model (others). In the red curve, the runaway current density is inferred by integrating the phase space current density above $p_V$. The blue curve assumes $v_{RE}=c$. The magenta and green markers apply test particle method (TPM) to describe the evolution of RE momentum with the different initial momentum $p_V$ and $1.3 p_c$, respectively. The black curve only consider an electric acceleration and defines the runaway boundary as $p_c$. The violet marker is obtained by using the reduced model.}
\label{fig:jRE_verf}
\end{figure}
To verify the multi-fluid RE current model, we borrow the test problem set in Fig. \ref{fig:param_evol}. Runaway avalanche and transport were neglected to focus on the dominant Dreicer generation. Figure \ref{fig:jRE_verf} shows time evolution of runaway current density. The red dashed curve in Fig. \ref{fig:jRE_verf} corresponds to the red dashed curve in Fig. \ref{fig:nRE_full}. Despite adequacy of the single fluid RE density estimate (see agreement with the magenta marker in Fig. \ref{fig:nRE_full}), the runaway current density (blue curve in Fig. \ref{fig:jRE_verf}) is overestimated within an order of magnitude under the approximation $v_{RE}=c$. However, the multi-fluid mildly relativistic correction significantly alleviates an error by relaxing the condition $v_{RE}=c$, as shown in the magenta marker. In real situation, it's difficult to know the correct $p_V$ without dedicated analysis. Instead of that, $p_{V} \approx 1.3 \ p_{c}$ (or another tuned coefficient) can be a good alternative for practical implementation.

\subsection{Single fluid approximation of mildly relativistic runaway current\label{ssec:SFM}}
Since the number of fluid runaway species gradually increases as simulation prolongs and raises the computational cost, we reduce the multi-fluid model into the single fluid form by assuming that the mean runaway velocity well characterizes the overall runaway current evolution. The mean RE velocity is given by $\beta_{RE,i} \equiv \frac{j_{RE,i}}{ecn_{RE,i}}$ and accordingly the single fluid model reads
\begin{equation}
\begin{split}
    \Big(\frac{dn_{RE}}{dt}\Big)_i &= n_{RE,i} (\gamma_{ava,i}-\frac{1}{\tau_{RE,i}}) + S_{p,i} \\
    \Big(\frac{dj_{RE}}{dt}\Big)_i &= ec\Big( n_{RE,i} (\gamma_{ava,i}-\frac{1}{\tau_{RE,i}}) \beta_{RE,i} \\
    &+ S_{p,i} \beta_{V,i}\Big) + ecn_{RE,i} \Big(\frac{d\beta_{RE}}{dt}\Big)_i,
\end{split}
\end{equation}
where $\beta_V = \frac{p_V}{\sqrt{p_V^2+1}}$ accounts for the seed velocity with the approximation $p_V \approx 1.3 p_c$ \footnote{The 1.3 factor is taken from the KSTAR-relevant test problem, and this choice need not necessarily be followed for other devices.} and $\frac{d\beta_{RE}}{dt}$ is described by Eq. \ref{eq:TPM}. The mean velocity resolves two important phases: 1) when the seed RE is strongly being created, it should represent the newly born RE particle velocity and 2) when the seed RE is barely created, it should represent the pre-existing RE velocity. Indeed, for sufficiently small $\Delta t$ with $n_{RE,i+1}^{I} =S_p \Delta t$, $n_{RE,i+1}^{II} = n_{RE,i}(1+(\gamma_{ava,i}-1/\tau_{RE,i}) \Delta t)$ and $\beta_{RE,i}^{\Delta t} = \beta_{RE,i}+(\frac{d\beta_{RE}}{dt})_i \Delta t$, the mean velocity at the $i+1$-th step reduces to 
\begin{equation}
\begin{split}
    \beta_{RE,i+1} &\approx \frac{n_{RE,i}^{II} \beta_{RE,i}^{\Delta t} + n_{RE,i}^I \beta_V}{n_{RE,i}^{II} + n_{RE,i}^{I}} \\
    &= 
    \begin{cases}
        \beta_{RE,i}^{\Delta t} \quad \text{if $n_{RE,i}^{II} \gg n_{RE,i}^I$} \\ 
        \beta_{V} \quad \quad \text{if $n_{RE,i}^{II} \ll n_{RE,i}^I$}
    \end{cases} \label{eq:bRE_single}
\end{split}
\end{equation}
where $n_{RE,i+1}^{I}$ and $n_{RE,i+1}^{II}$ represent the newly born RE seed during $\Delta t$ and pre-existing RE beam before $t_i$, respectively. When two populations are comparable, the interpolating formula \ref{eq:bRE_single} represents an intermediate velocity. We identified that the accuracy of the single fluid model (violet diamond marker) is comparable to free fall calculation (black dotted curve) in Fig. \ref{fig:jRE_verf}.

\subsection{Effective critical electric field}
Significant amount of neutral particles of main fuel gas and impurities are present during the burn-through phase. Low collisionality allows the critical RE energy to be close to the ionization potential, in which the inelastic interactions predominantly involve large energy transfers. In such a system, the friction-only treatment of inelastic collisions can underestimate the RE generation rate significantly \cite{Lee2024PRL}.

As an ad-hoc treatment of including inelastic collisions in the critical electric field, we introduce the effective critical field \cite{Martinsolis2015PoP} where the "soft" inelastic collisions are additionally considered to account for the binary nature \cite{Lee2024PRL,Lee2026JPP}:
\begin{equation}
\begin{split}
    &E_c \equiv \frac{e^3 n_e \ln \Lambda_{free}}{4\pi \varepsilon_0^2 mc^2} \to \\
    &E_c^{eff} \equiv \frac{e^3 (n_e \ln \Lambda_{free}+n_H \ln\Lambda_{bound}^{H,soft}+\sum_I n_I \ln\Lambda_{bound}^{I})}{4\pi \varepsilon_0^2 mc^2} \label{eq:Ec}
\end{split}
\end{equation}
where the $H$ denotes the main neutral species (hydrogen or deuterium or tritium) and $I$ is the impurity species. The treatment effectively increases the critical momentum in phase space. For impurities, we assume dominant soft inelastic collisions because $p_c$ is high enough to neglect the ionization potential when a strong impurity influx is expected \cite{Lee2023PS, Lee2023NF}.

\subsection{Effective Dreicer generation model}
As an ad-hoc treatment of including inelastic collisions in the Dreicer generation model, we evaluate the Connor-Hastie formula \cite{Connor1975NF} by using the effective Dreicer field \cite{Martin-solis2017NF}
\begin{equation}
    E_d \equiv E_c \frac{mc^2}{T_e}  \to E_d^{eff} \equiv E_c^{eff} \frac{mc^2}{T_e}. \label{eq:Ed}
\end{equation}
Note that this treatment only consider the modification in $p_c$ with the binary nature of inelastic collisions, but neglect the inelastic energy diffusion. Another similar ad-hoc treatment to include the inelastic energy diffusion was proposed by Ref. \cite{Gurevich1961JETP}. But we don't consider this herein since inelastic energy diffusion has a strong energy-dependence and such an ad-hoc estimation likely leads to overestimation \cite{Lee2026arXiv}.

\subsection{Runaway avalanche model}
We adopt the Rosenbluth-Putvinski formula \cite{Rosenbluth1997NF} with $E_c^{eff}$ \cite{Martinsolis2015PoP} to describe the runaway avalanche model. This simplification is justified for the following reasons. (1) Hard ionizing collisions felt by seed REs can drive collisional decay (reverse of amplification) but its decay rate is expected to be weaker than runaway deconfinement rate during early startup, i.e. $\gamma_{ava}-\frac{1}{\tau_{RE}} \approx -\frac{1}{\tau_{RE}}$. (2) When the seed population is dominant after CFSF, possible errors in the growth rate arising from the finite RE energy correction are not significant, i.e. $S_p + n_{RE}\gamma_{ava} \approx S_p$ \cite{Lee2023NF}. (3) In KSTAR, a range of $E/E_c$ is far higher than the so-called near critical regime \cite{Aleynikov2015PRL} and the infinite RE energy assumption used in Ref. \cite{Rosenbluth1997NF} can hold for REs with $p \geq 2.3$ within $5 \%$ error (See Figure 9 in Ref. \cite{Lee2023NF}).

\section{Self-consistent coupling between the mildly relativistic runaway electron model and full electromagnetic plasma initiation model in DYON \label{sec:coupling}}
We self-consistently couple the runaway current circuit with the full circuit equation and describe the runaway transport with the model-based time evolving magnetic configurations in DYON \cite{Kim2020NF}. This predictive description enables the startup RE assessment for future device designs and operation scenarios.

In DYON-RE, the electric field can be either a local electric field calculated from the time evolution of two-dimensional magnetic flux or a global electric field. We use the resistive electric field in this work (see Sec. \ref{ssec:inhomogeneity}), where only the burn-through and early rampup phases are modelled.

\subsection{Effective runaway transport under dual magnetic configurations \label{ssec:RE_conf_model}}
In the plasma burn-through phase during tokamak startup, magnetic configurations coexist with each other. Meanwhile, the RE transport is highly sensitive to whether the configuration is open or closed. Therefore, there are two macroscopic RE populations, $n_{RE}^{cl}$ and $n_{RE}^{op}$, within the closed flux surface and in open field lines, respectively. Each populations are described by the single fluid treatment, not the multi fluid. The particle-conserving evolution of RE density in each configuration is given by
\begin{equation}
    \frac{dn_{RE}^{cl}}{dt} = S_p^{cl} + (\gamma_{ava}^{cl,i} -  \frac{1}{\tau_{RE,\perp}^{cl}}) n_{RE}^{cl} - n_{RE}^{cl} \frac{d}{dt} \log V_p^{cl}, \label{eq:dndt_cl}
\end{equation}
\begin{equation}
\begin{split}
    \frac{dn_{RE}^{op}}{dt} &= S_p^{op} + (\gamma_{ava}^{op,i} -  \frac{1}{\tau_{RE}^{op}}) n_{RE}^{op} + \frac{V_p^{cl}}{V_p^{op}} \frac{n_{RE}^{cl}}{\tau_{RE,\perp}^{cl}} \\
    &- n_{RE}^{op} \frac{d}{dt} \log V_p^{op}. \label{eq:dndt_op}
\end{split}
\end{equation}
The terms containing $\frac{d}{dt} \ln V_p^{op}$ and $\frac{d}{dt} \ln V_p^{cl}$ describe the density evolution arising from changes in the open and closed region volumes, respectively. The third term in RHS of Eq. \ref{eq:dndt_op} $\frac{V_p^{cl}}{V_p^{op}} \frac{n_{RE}^{cl}}{\tau_{RE,\perp}^{cl}}$ represents the particle source transported from the closed region, which is required for the particle conservation within plasmas but plays a minor role in runaway dynamics in the open field region. Similarly, the momentum-conserving evolution of RE current density in each configuration is written as
\begin{equation}
\begin{split}
    \frac{dj_{RE}^{cl}}{dt} &= ecS_p^{cl} \beta_V^{cl} + (\gamma_{ava}^{cl} -  \frac{1}{\tau_{RE,\perp}^{cl}}) j_{RE}^{cl} + j_{RE}^{cl} \frac{d}{dt} \log \beta_{RE}^{cl} \\
    &- j_{RE}^{cl} \frac{d}{dt} \log V_p^{cl} ,\label{eq:djdt_cl}
\end{split}
\end{equation}
\begin{equation}
\begin{split}
    \frac{dj_{RE}^{op}}{dt} &= ecS_p^{op} \beta_V^{op} + (\gamma_{ava}^{op} -  \frac{1}{\tau_{RE}^{op}}) j_{RE}^{op}+ j_{RE}^{op} \frac{d}{dt} \log \beta_{RE}^{op} \\
    &+ \frac{V_p^{cl}}{V_p^{op}} \frac{j_{RE}^{cl}}{\tau_{RE,\perp}^{cl}} - j_{RE}^{op} \frac{d}{dt} \log V_p^{op}.\label{eq:djdt_op}
\end{split}
\end{equation}
Provided that $\frac{dI}{dt} = \frac{d}{dt}(j\cdot A)$, Eqs. \ref{eq:djdt_cl} and \ref{eq:djdt_op} are converted into the runaway current forms,
\begin{equation}
\begin{split}
    \frac{dI_{RE}^{cl}}{dt} &=  ecA^{cl}S_p^{cl} \beta_V^{cl}+ I_{RE}^{cl} \frac{d}{dt} \log \beta_{RE}^{cl} \\
    &+ (\gamma_{ava}^{cl,j} -  \frac{1}{\tau_{RE}^{cl}}) I_{RE}^{cl} ,\label{eq:dIdt_cl}
\end{split}
\end{equation}
\begin{equation}
\begin{split}
    \frac{dI_{RE}^{op}}{dt} &= ecA^{op}S_p^{op} \beta_V^{op} + (\gamma_{ava}^{op} -  \frac{1}{\tau_{RE}^{op}}) I_{RE}^{op} \\
    &+ I_{RE}^{op} \frac{d}{dt} \log \beta_{RE}^{op} + \frac{R^{cl}}{R^{op}} \frac{I_{RE}^{cl}}{\tau_{RE}^{cl}}\label{eq:dIdt_op}
\end{split}
\end{equation}
where $j_{RE}^{cl} \frac{d}{dt} \log V_p^{cl}$ and $j_{RE}^{op} \frac{d}{dt} \log V_p^{op}$ approximately cancel with $j_{RE}^{cl} \frac{d}{dt} \log A_p^{cl}$ and $j_{RE}^{op} \frac{d}{dt} \log A_p^{op}$, respectively.

The RE transport mechanisms differ according to magnetic configurations. In closed field configuration, REs only have the radial transport. The radial transport model should be carefully chosen with the model assumptions in consideration since the RE density evolution is sensitive to the RE confinement time after the seed formation phase as shown in Ref. \cite{deVries2025NF}. Because the RE energy is insufficient for a large orbit shift at the early phase \cite{Knoepfel1979NF, Vries2023NF}, we neglect the drift-orbit loss and only consider the Rechester-Rosenbluth diffusion \cite{Rechester1978PRL},
\begin{equation}
    \tau_{RE}^{cl} \approx \tau_{RE,\perp}^{cl} \approx c_{RR}\frac{a_r^2}{2\pi R_0 v_{RE}^{cl}} (\frac{\tilde{b}_r}{B_\phi})^{-2} \label{eq:tauRE_cl},
\end{equation}
where $c_{RR}$ is a free parameter effectively measuring a correlation radial length and $\frac{\tilde{b}_r}{B_\phi}$ is the normalized radial magnetic fluctuation. We choose $c_{RR}|\frac{\tilde{b}_r}{B_\phi}|^{-2}\approx 1.25\times10^{9}$ to phenomenologically explain the ECE behaviors in KSTAR (see Sec. \ref{ssec:DYON-RE}). 
In open field configuration, the parallel transport is taken into account,
\begin{equation}
    \tau_{RE}^{op} \approx \tau_{RE,\parallel}^{op}\approx <L_{op}> / v_{RE}^{op} \label{eq:tauRE_op}
\end{equation}
where $\tau_{RE,\perp}^{cl}$ and $\tau_{RE,\parallel}^{op}$ are descried by  and parallel streaming loss, respectively.

\subsection{Amended circuit system}
The amended circuit equation is
\begin{equation}
    \frac{dI_p}{dt} = \frac{1}{L_p} \Big( V_{loop} - R_p (I_p - I_{RE}^{op} - I_{RE}^{cl}) - \frac{1}{2} \frac{dL_p}{dt} I_p \Big), \label{eq:dIpdt}
\end{equation}
which extends Eq. (6) of Ref. \cite{Kim2022NF} to include the runaway current; $I_p$ is the total current. The amendment is to consider the resistive voltage ($V_{res}=R_p (I_p - I_{RE}^{op} - I_{RE}^{cl})$) by subtracting the total runaway current. In Ref. \cite{Kim2022NF}, the local electric field is correctly described through the magnetic flux conservation when determining the Townsend avalanche onset, and the circuit equation is accordingly formulated without a $\frac{1}{2}$ factor in front of $\frac{dL_p}{dt}$. By contrast, we retain the $\frac{1}{2}$ factor to remain consistent with the energy conservation description \cite{Ejima1982NF} discussed in Sec. \ref{ssec:RPBT}. We couple this with the circuit system of DYON \cite{Kim2022NF}. In the energy-conserving description, rigorously speaking, $V_{loop}$ in Eq. \ref{eq:dIpdt} is the loop voltage measured at the conducting boundary (or that measured at the inboard wall) and different from that measured at the plasma center used in DYON \cite{Kim2022NF}. Yet, we used the loop voltage at the plasma center following the DYON model convention. This approximation is not valid when an inhomogeneity in an electric field is significant. In that case, a local electric field should be considered.

In a superconducting tokamak, the time scale of runaway current development in open field configuration is much faster than the resistive time scale of total current. This facilitates the dominant balancing
\begin{equation}
    n_{RE}^{op} \approx S_p^{op}\tau_{RE}^{op},
\end{equation}
\begin{equation}
    I_{RE}^{op} \approx ecA^{op}S_p^{op} \beta_V^{op}\tau_{RE}^{op},
\end{equation}
with an amended resistive electric field ($=R_p(I_p - I_{RE}^{op} - I_{RE}^{cl})/2\pi R_0)$ iteratively found.

\begin{figure*}[ht!]
    \centering
    \includegraphics[width=\textwidth]{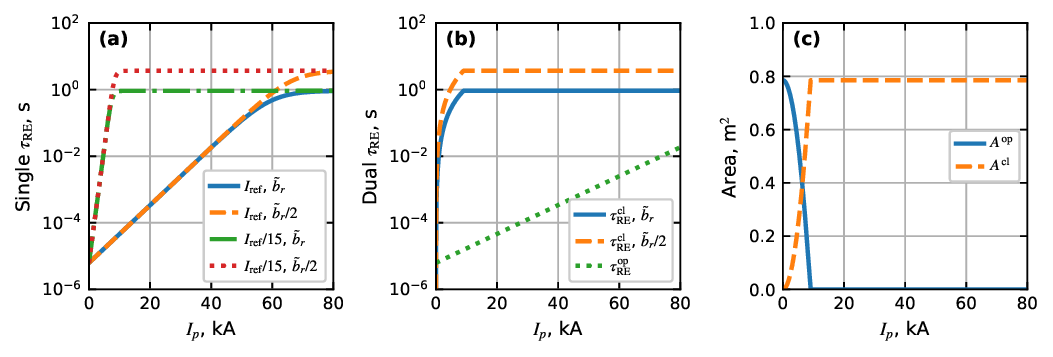}
    \caption{$\tau_{RE}$ of the single population (a) and dual populations (b) and plasma cross sectional area (c) as a function of $I_p$. In (a,b) legend, $\tilde{b}_r$ corresponds to $c_{RR}|\frac{\tilde{b}_r}{B_\phi}|^{-2}\approx 1.25\times10^{9}$ and $\tilde{b}_r/2$ does to $c_{RR}|\frac{\tilde{b}_r}{B_\phi}|^{-2}\approx 5\times10^{9}$. In (a), the blue and orange curves use $I_{RE,ref}=I_{ref}$ and the green and red curves use $I_{RE,ref}=I_{ref}/15$.}
    \label{fig:RE_confinement}
\end{figure*}

\subsection{Effect of local closed flux surface formation on seed runaway confinement \label{ssec:LCFSF}}
The \textit{global} closed flux surface refers to the fully developed state in which nearly all field lines have closed and no plasmas reside on open field lines. By contrast, the \textit{local} closed flux surface refers to a developing state in which closed flux surfaces coexist with plasmas still on open field lines. The dual magnetic configurations introduced in Sec. \ref{ssec:RE_conf_model} were conceived to isolate the radial transport of runaway electrons inside the local closed flux surface from the parallel transport in the open field region. This section examines how this description affects the runaway confinement time before the global CFSF, while the resulting impact on runaway dynamics will be addressed at Sec. \ref{ssec:noEM}.

Unlike the single population description, the dual population consideration relies on the open and closed regions described by DYON-EM. To compare two models fairly by removing the model dependence, we prescribe a simplified plasma geometry which has a fixed plasma volume but separates the open and closed regions depending on the magnitude of plasma current. Plasma parameters are chosen to be relevant to the KSTAR startup: the plasma region has a circular shape with the standard minor radius $a=0.5 \, m$. The RE velocities are assumed to be $\beta_{RE}=\beta_{RE}^{op}=\beta_{RE}^{cl}=0.1$. We infer the effective connection length using \cite{Lloyd1991NF, Kim2012NF}
\begin{equation}
    L_{eff} = 0.25 a \Big(\frac{B_\perp}{B_\phi}\Big) \exp(\frac{I_p}{I_{ref}}).
\end{equation}
The RE parallel confinement time is
\begin{equation}
    \tau_{RE,\|}=\tau_{RE}^{op} \approx \frac{L_{eff}}{v_{RE}}.
\end{equation}
In KSTAR, the global closed flux surface forms when a plasma current is above $10 \, kA$ \cite{Lee2023PS}. Hence, we estimate a stray magnetic field level $\frac{B_\perp}{B_\phi}\approx 0.2 \, \%$ and a reference plasma current for the CFSF $I_{ref}=10 \, kA$. The closed region is then prescribed as a region where the plasma-induced magnetic field exceeds the stray field $B_\perp$ with the assumption that the plasma current is concentrated on a magnetic axis as usually adopted in the DYON electromagnetic modelling \cite{Kim2022NF, Kim2024NF, Kim2026NF}; the open region is the remaining region. This provides a conservative prediction of the RE confinement. The associated minor radius is
\begin{equation}
    a_r \approx \min(\frac{\mu_0 I_p}{2\pi B_\perp}, a).
\end{equation}

The macroscopic RE confinement time can be found by using the interpolation proposed by Ref. \cite{Hoppe2022JPP} and subsequently adopted by Refs. \cite{Matsuyama2022NF, deVries2025NF}:
\begin{equation}
    \tau_{RE}^{-1} = \exp(-\frac{I_p}{I_{RE,ref}}) \tau_{RE,\|}^{-1} + (1-\exp(-\frac{I_p}{I_{RE,ref}})) \tau_{RE,\perp}^{-1} \label{eq:tau_RE_gloabal}
\end{equation}
where $\tau_{RE,\perp}$ is described by replacing $a_r$ in Eq. \ref{eq:tauRE_cl} with $a$, and $I_{RE,ref}$ is an interpolating coefficient.

Figure \ref{fig:RE_confinement} compares the macroscopic $\tau_{RE}$ (a) with the proposed dual $\tau_{RE}^{op}$ and $\tau_{RE}^{cl}$ (b) and shows the evolutions of $A^{op}$ and $A^{cl}$ (c). Through this comparison, we elucidate that although the proposed dual description does not produce any meaningful difference after the \textit{global} CFSF ($I_p > 10 \, kA$), $\tau_{RE}^{cl}$ is clearly longer than $\tau_{RE}$ during the \textit{local} CFSF ($I_p < 10 \, kA$). Indeed, the green and red curves in Fig. \ref{fig:RE_confinement}(a) indicate that $\tau_{RE}$ is saturated to $\tau_{RE,\perp}$ after the global CFSF ($I_p > I_{ref}$) when $I_{RE,ref}=I_{ref}/15$ is chosen. This suggests the interpolating form (Eq. \ref{eq:tau_RE_gloabal}) is adequate to capture the RE transport after the global CFSF so applicable to the early rampup RE modelling considered in Refs. \cite{Hoppe2022JPP, deVries2025NF}. However, the transport during the \textit{local} CFSF needs an additional resolution. Our RE transport model describes the improved confinement by showing that $\tau_{RE}^{cl}$ represented by the orange and blue curves in Fig. \ref{fig:RE_confinement}(b) is much higher than $\tau_{RE}$ represented by the red and green curves in Fig. \ref{fig:RE_confinement}(a) before the global CFSF. Hence, we conclude that when the startup RE seed \footnote{The CFSF proceeds together with the burn-through, and as the discharge transitions into the early current rampup, the plasma temperature rises and the resistive electric field decreases, weakening the primary generation. The REs formed through this "first" phase thus constitute a seed of the RE beam, which may subsequently grow via avalanche or decay via transport. A "second" period of enhanced primary generation may occur later during the early current rampup; in this context, "seed" refers to the REs formed in the first period described above.} is important it is desirable to account for the dual magnetic configurations whereas when it becomes negligible either description can be used without significant impact.

Note that using the macroscopic model requires a careful choice of $I_{RE,ref}$. As shown by Fig. \ref{fig:RE_confinement}(a), the direct evaluation $I_{RE,ref}=I_{ref}$ (orange and blue) can underestimate the $\tau_{RE}$ by several orders of magnitude compared to $\tau_{RE,\perp}$ after the global CFSF. From this perspective, it is worth checking that $\tau_{RE}$ adequately represents $\tau_{RE,\perp}$ before significant primary generation begins during the early current rampup.

Recall that in this section we intentionally prescribed the stray field and plasma area for the model comparison. We replace these simplifying assumptions with the DYON electromagnetic model for the KSTAR application.

\section{Qualitative comparison of DYON-RE in KSTAR ohmic startup \label{sec:validation}}
\subsection{Observation of startup runaway electrons in KSTAR Ohmic discharges}
Ohmic discharges in 2020 KSTAR campaign were primarily used as the reference shots for comparison of RE model. The formation of runaway electrons was observed in the reference discharges \cite{Lee2022EPS}. These shots share a very narrow prefill window, while engineering parameters in the control room is similar. It leads to the expectation that a similar amount of runaway electron generation would have occurred across shots. However, we observed pronounced differences in runaway electron behavior between discharges.

\begin{figure}[h!]
    \centering
    \includegraphics[width=0.5\textwidth]{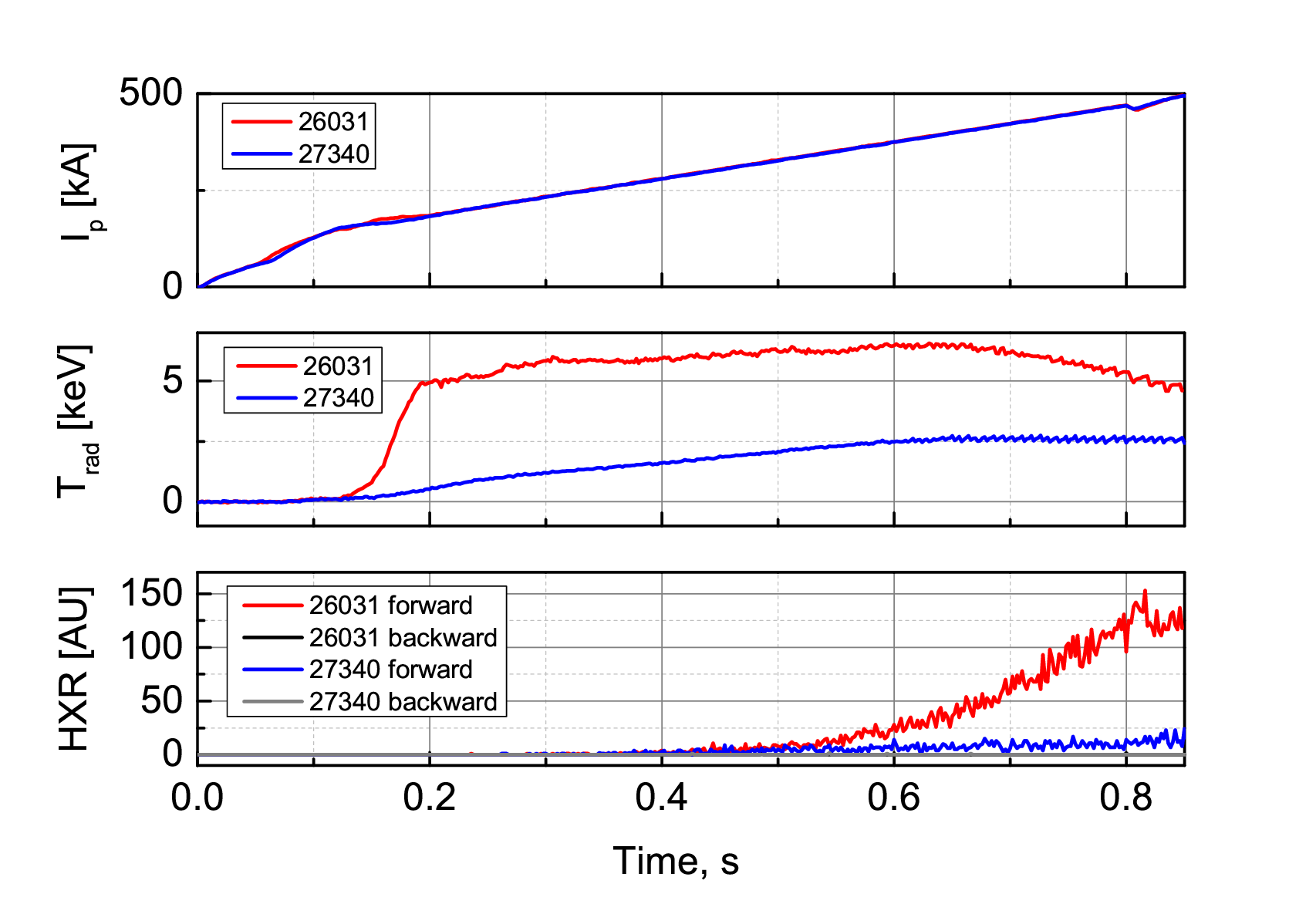}
    \caption{Comparison of RE signs between RE rich shot ($\#$26031, red) and RE scarce shot ($\#$27340, blue). Top panel is $I_p$ in $keV$, middle panel is the core $T_{rad}$ in $eV$ measured at $R=1.8$ m and bottom panel is the forward HXR intensity in $AU$. Grey and black curves in the bottom indicate the absence of the backward HXR signal.}
    \label{fig:comparison}
\end{figure}

KSTAR is equipped with Electron Cyclotron Emission (ECE) Radiometer \cite{Lee2022JOI} assigned to the K port located at low field side and Hard X-Ray (HXR) diagnostics based on NaI(Tl) scintillator \cite{Lee2011FST}, both of which are main diagnostics indirectly indicating signs of REs. Roughly speaking, the non-thermal ECE reflects the presence of REs at the sub-MeV (hundreds of keV) level, whereas the HXR signal indicates the loss of more energetic, MeV-scale REs.

In an optically thin Maxwellian medium, the radiation intensity does not reach the black-body intensity and the radiative temperature is lower than the electron temperature. Accordingly, we attribute the rising ECE intensity to the non-thermal effect, which in turn implies the RE formation.

HXR signal implies RE loss due to collision with the limiter. HXR intensity corresponds to the number of photons aggregated over $1 \, \mathrm{ms} $ from 1024 channels, the number of which is proportional to energy of photons they detected but not calibrated. Counts of photon from channel 0 were regarded as a noise and thus excluded from the summation. The number of detected photons is insufficient for interpreting the photon energy distribution \cite{Paz-Soldan2018PoP}.

The discharges were classified as RE rich shots or RE scarce shots according to whether the maximum HXR count per $1 \, \mathrm{ms}$ during the rampup phase exceeded 100. Figure \ref{fig:comparison} shows the different time evolutions of the $I_p$, $T_{rad}$ and forward HXR intensity of representative discharges $\#$26031 and $\#$27340 of the RE rich and RE scarce shots, respectively. $T_{rad}$ is measured by the ECE diagnostics of which frequency corresponds to the machine major radius ($R=1.8 \, \textrm{m}$). The discharges exhibit a similar $I_p$ waveform. However, $T_{rad}$ in $\#$26031 increased earlier than that in $\#$27340. In addition, overall $T_{rad}$ in $\#$26031 is far higher than that of $\#$27340, hinting more REs in $\#$26031. Also, once a photon produced by RE loss is detectable, the forward HXR intensity of $\#$26031 become higher than that of $\#$27340. This indicates a higher number of start REs survive in $\#$26031. It is noteworthy that HXR intensity does not follow the time trend of the ECE intensity in the figure \ref{fig:comparison}, since HXR intensity shows loss of highly energetic REs while ECE intensity shows creation of mildly energetic REs.

\begin{figure}[h!]
    \centering
    \includegraphics[width=0.5\textwidth]{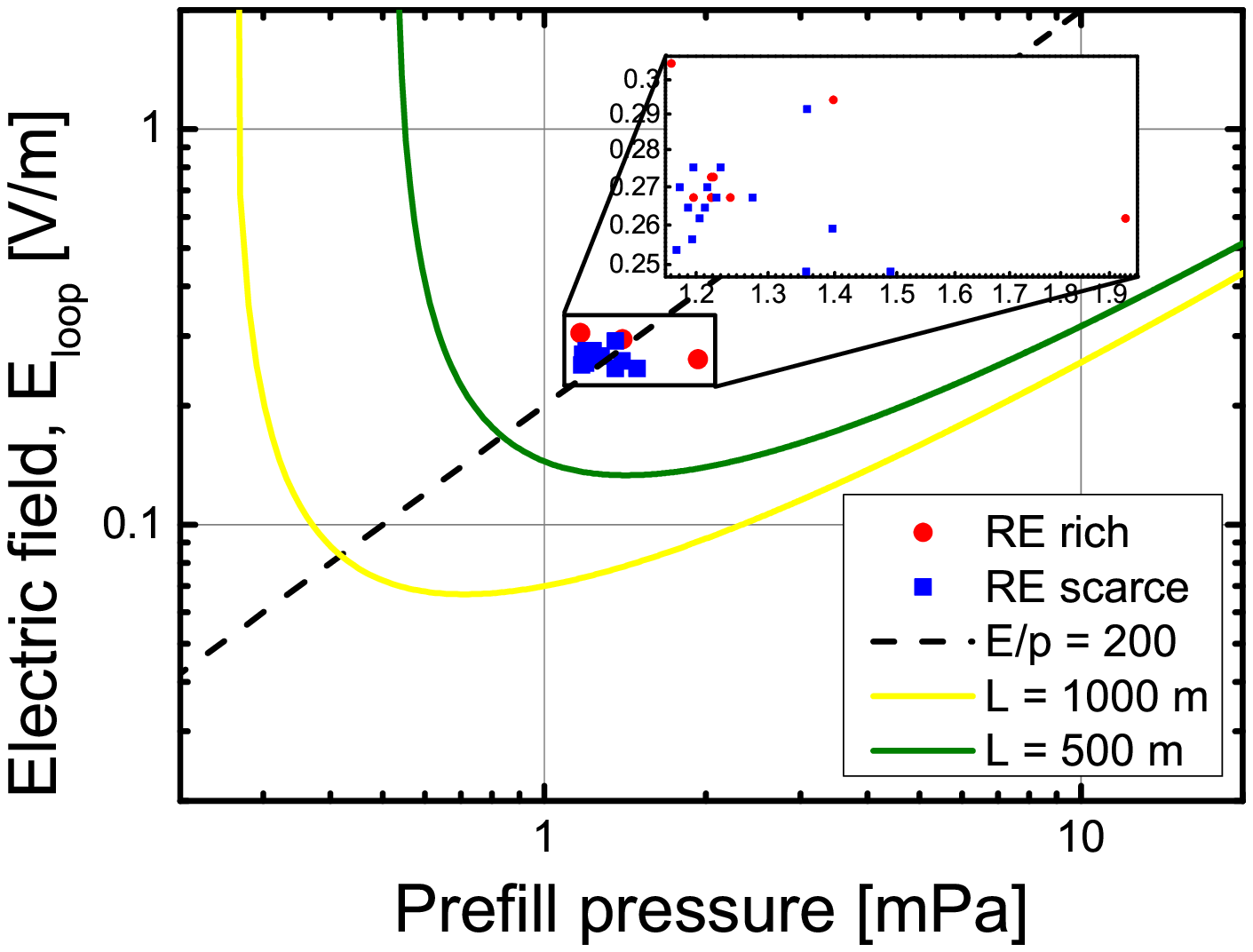}
    \caption{A scatter plot of the breakdown condition of the RE rich (red) and RE scarce discharges (blue). E/p has $Vm^{-1}Pa^{-1}$ unit. Yellow and green lines represents the minimum breakdown voltage with connection length, $L$.}
    \label{fig:BD_window}
\end{figure}

Figure \ref{fig:BD_window} shows the operational window of $E_{loop}$ and the prefill gas pressure. There are no significant differences in the initial discharge conditions between the RE-rich and RE-scarce groups.

\begin{figure}[h!]
    \centering
    \includegraphics[width=0.5\textwidth]{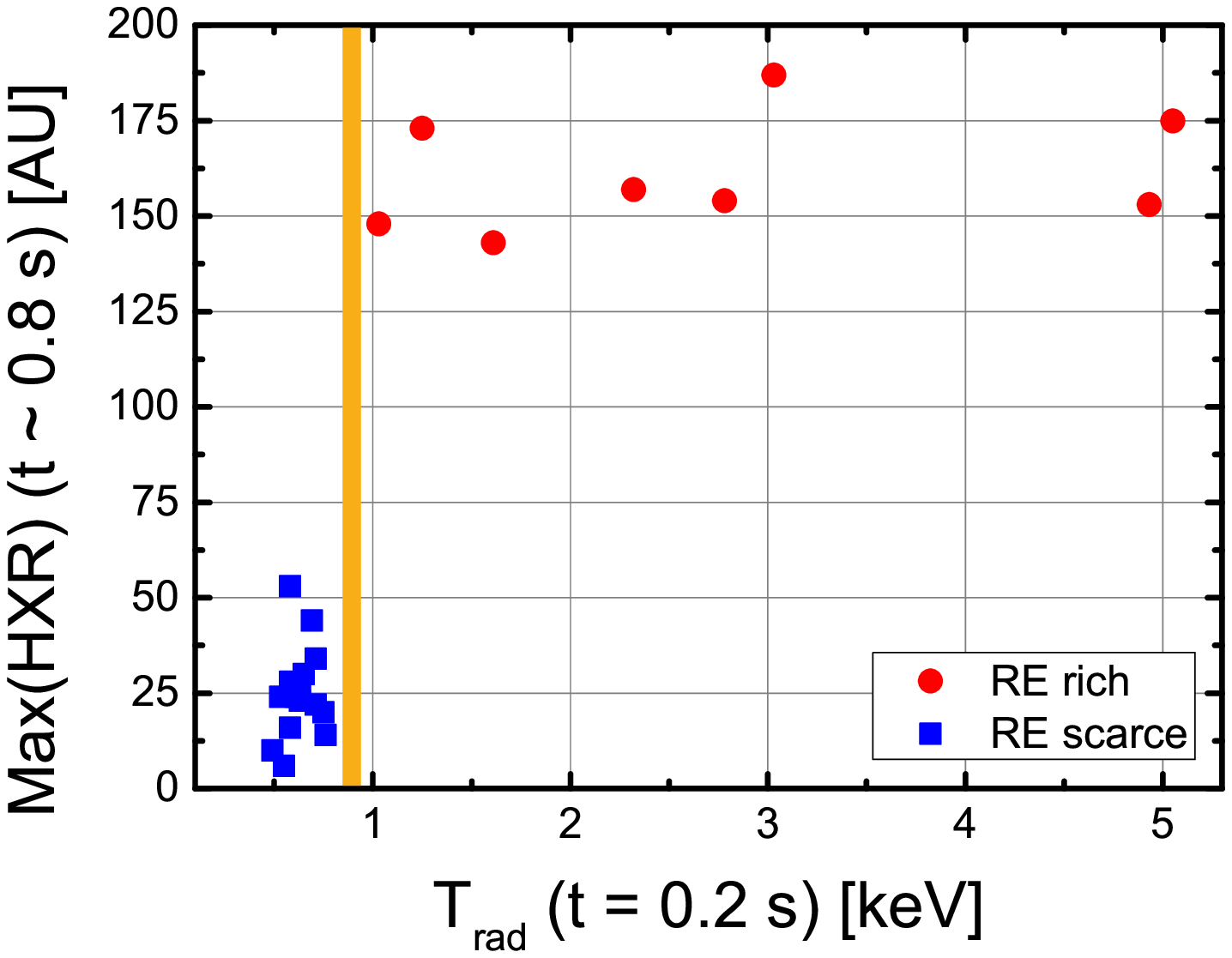}
    \caption{(a) A scatter plot of the maximum HXR intensity during rampup against the radiative temperature measured by the core ECE corresponding to $R=1.8 \ m$ at t = 0.2 s of the RE rich (red) and RE scarce discharges (blue).}
    \label{fig:Bif}
\end{figure}
In contrast, as suggested by Fig. \ref{fig:Bif}, a clear distinction between the two groups emerges when comparing the level of early non-thermal ECE signals measured by the core ECE channel. This indicates that the observed differences in runaway electron populations did not arise from a specific event during the rampup phase, but rather originated during the initial startup phase. These early-formed runaway electrons persist throughout the rampup and gain additional energy, ultimately leading to the differences observed in the HXR signals.

\begin{figure}[h!]
    \centering
    \includegraphics[width=0.5\textwidth]{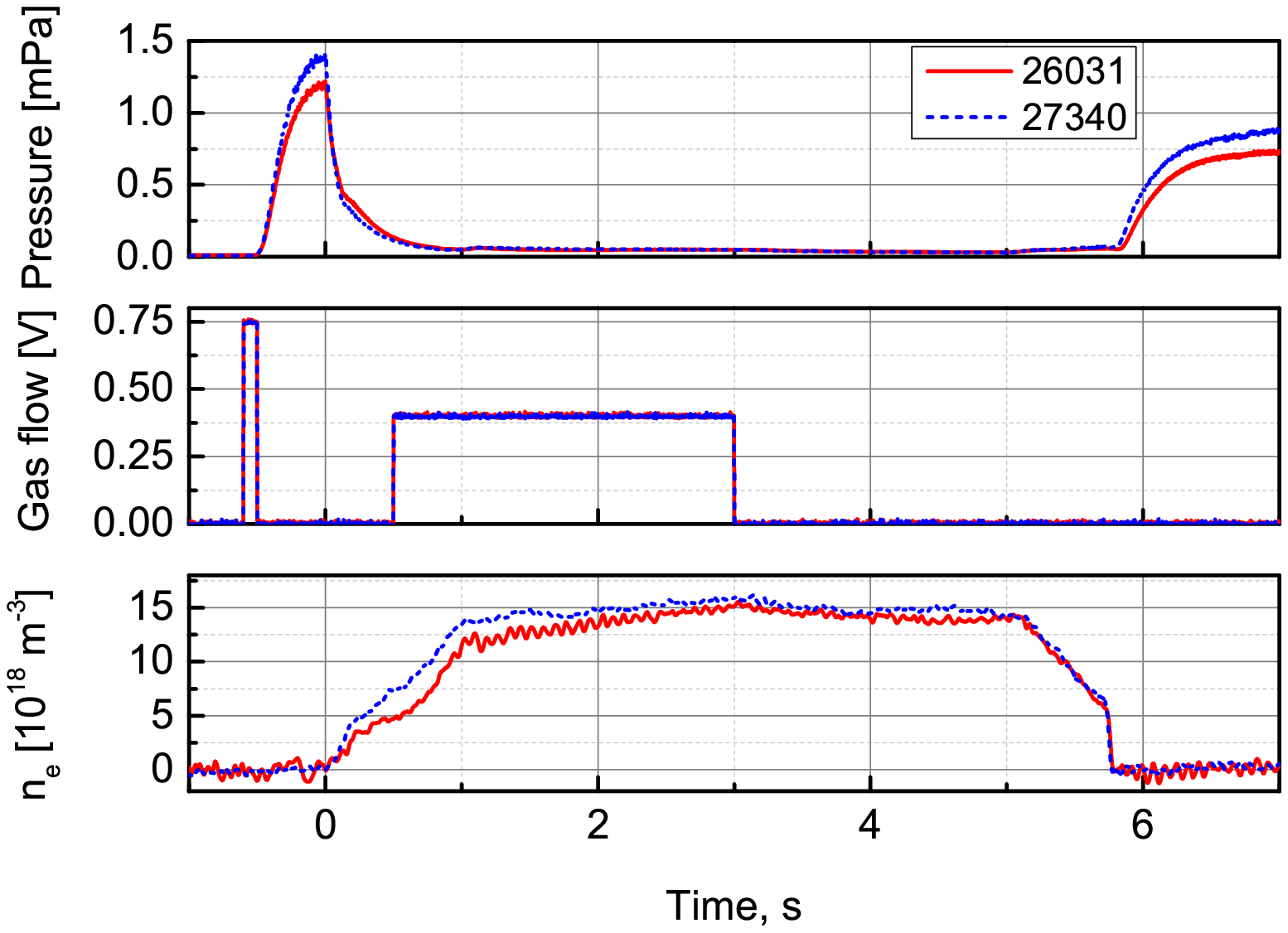}
    \caption{Time evolutions of neutral pressure in mPa (top panel), gas flow voltage in V (middle panel) and line-averaged density $n_e$ in $10^{18}$ $m^{-3}$ (bottom panel). Red-solid curve is $\#$26031 and blue-dashed curve is $\#$27340.}
    \label{fig:gflow}
\end{figure}
According to the experimental observations, one possible mechanism that can account for this trend is a significant difference in wall conditions between discharges. Evidence supporting this interpretation is shown in Fig. \ref{fig:gflow}. The gas flow data in the middle panel suggests that the gas valve voltage applied to the system was identical across the discharges. However, the prefill gas pressure is slightly higher in $\#27340$ (top panel). Also, as the plasma terminates, the plasmas neutralizes and the corresponding neutral pressure exhibits a stronger magnitude. Indeed, the electron density remains higher in $\#27340$ (bottom panel), consistent with the trend in the neutral pressure evolutions.

Based on these results, we conclude that the wall conditions were not identical between the two discharges, and it should be the main driver of the different REs.

\begin{figure}[h!]
    \centering
    \includegraphics[width=0.4\textwidth]{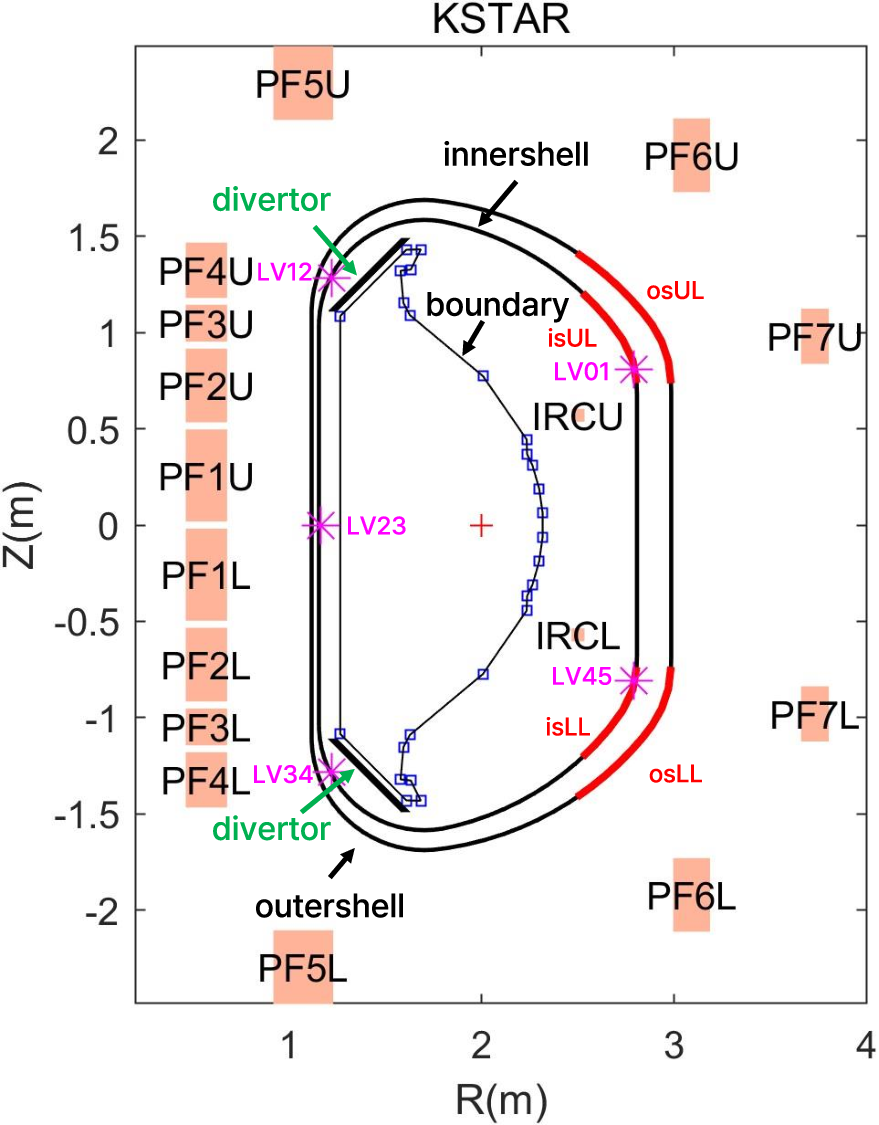}
    \caption{Coils and passive structures consisting of divertor, innershell and outershell in KSTAR. Only resitivities for inner/outer upper/lower structures at LFS (red lines marked as isUL, osUL, isLL and osLL) and divertors are effectively considered. In the Townsend calculation, a plasma boundary is simplified as black solid curve with blue square markers. Locations of loop voltage channels (01, 12, 23, 34 and 45) are also illustrated. PF is the poloidal field coil and IRC is the in-vessel radial control coil. U and L denote the upper and lower parts, respectively.}
    \label{fig:vacuum_geometry}
\end{figure}
\subsection{Vaccum vessel model adjustment}
The DYON-EM simulation takes toroidal eddy currents flowing through passive structures into consideration using the circular coil approximation, in which a three-dimensional geometry of passive structure in a reality is simplified as two-dimensional (2D). Figure \ref{fig:vacuum_geometry} shows the two-dimensional KSTAR geometry with coils, passive structures and plasma boundary. Modelled passive structures comprise innershell, outershell and divertors. Inside the vessel, there are also other components \cite{Bak2023FED} but we only consider the divertors in the model. To justify the 2D simplification, we adjusted the vacuum vessel model, in which the resistances in some passive structures are calculated with effective resistivities to accurately reproduce the electromagnetic responses in the plasmaless (vacuum) discharge. The responses we aimed to benchmark against are (1) loop voltages measured by five flux loops (see magenta $x$-markers, 01, 12, 23, 34 and 45), (2) total toroidal eddy currents measured by a rogoski coil inside innershell and (3) stray field profiles reconstructed from the FIST99 code \cite{Kim2011NF}. 
(1) and (3) guarantee a reliable representation of the electromagnetic field structure within the plasma on which DYON-EM performs the Townsend assessment to estimate the plasma volume. In addition, (2) enables a clear discrimination of the plasma-current fraction in the net current measured by the Rogowski coil. Hence, our targets (1)-(3) are sufficient for the purpose of comparison.

\begin{figure}[h!]
    \centering
    \includegraphics[width=0.5\textwidth]{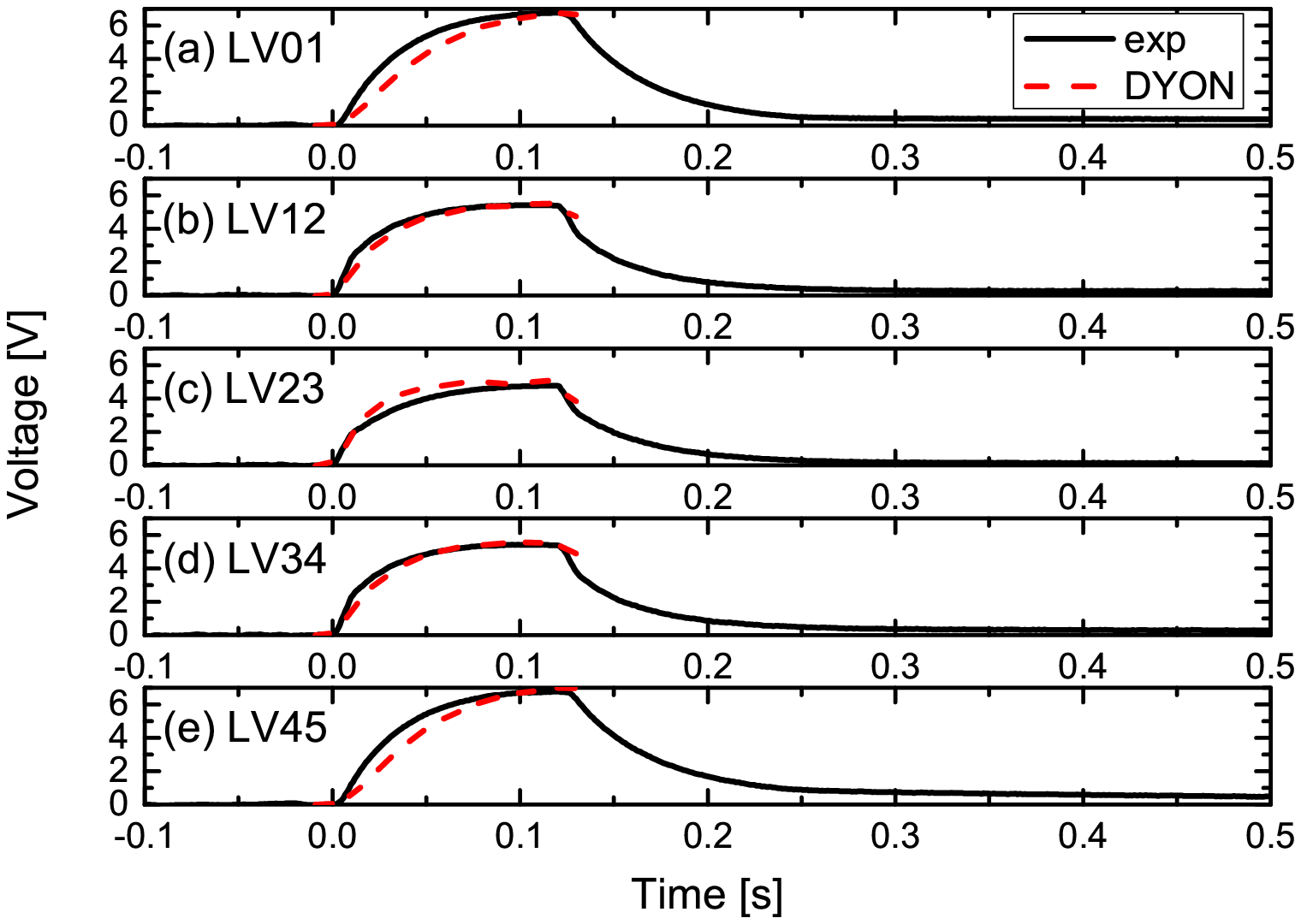}
    \caption{Comparison of loop voltages between experiment (black solid) and DYON predictions (red dashed). Locations of each channel can be found in Fig. \ref{fig:vacuum_geometry}.}
    \label{fig:vacuum_lv}
\end{figure}

\begin{figure}[h!]
    \centering
    \includegraphics[width=0.5\textwidth]{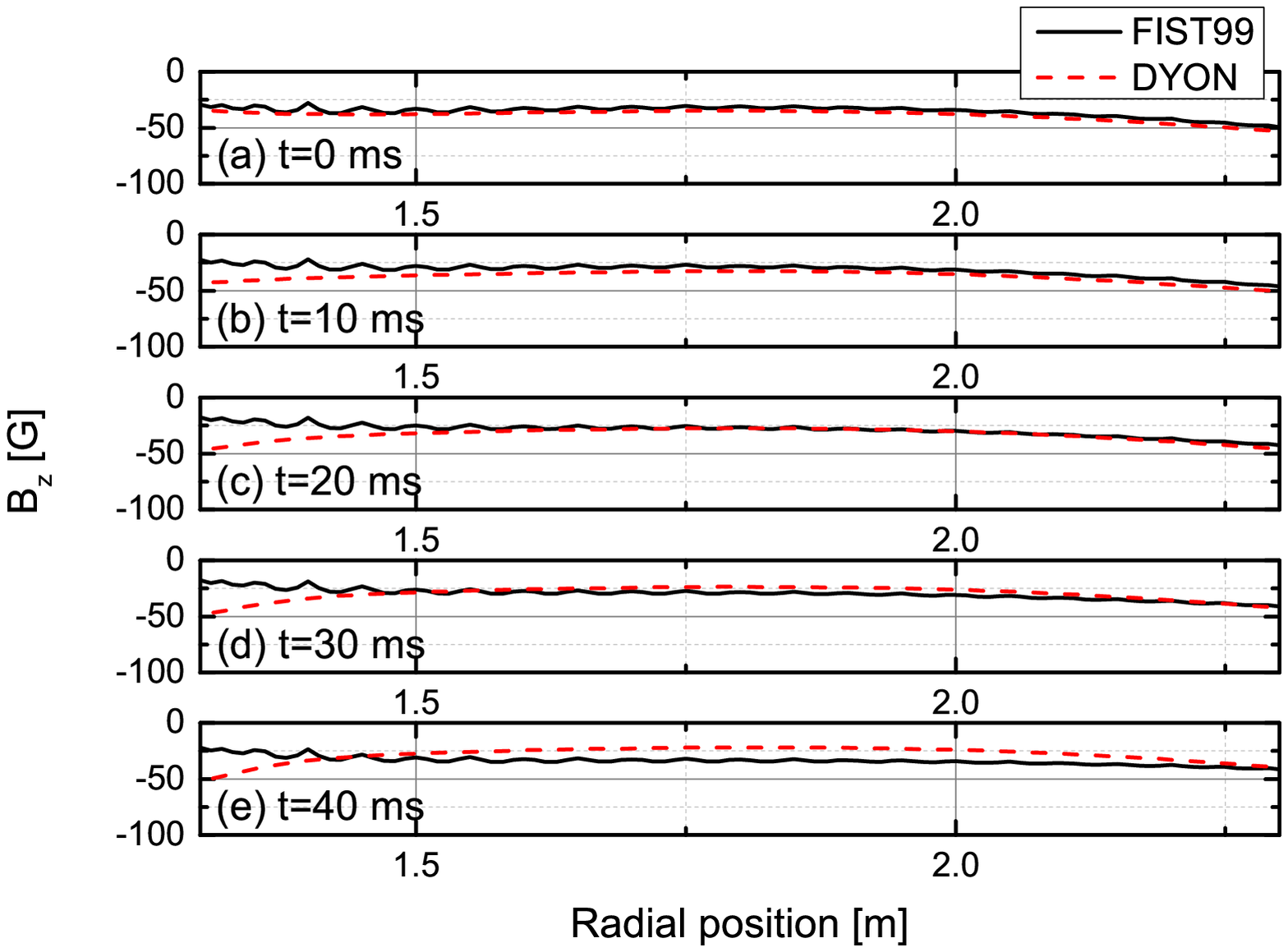}
    \caption{Comparison of $B_z$ in G ($= 0.1 \, \mathrm{mT}$ in the SI unit)  between FIST99 prediction (black solid) and DYON predictions (red dashed). $Z = 0 \ m$ is selected.}
    \label{fig:vacuum_null}
\end{figure}

\begin{figure}[h!]
    \centering
    \includegraphics[width=0.5\textwidth]{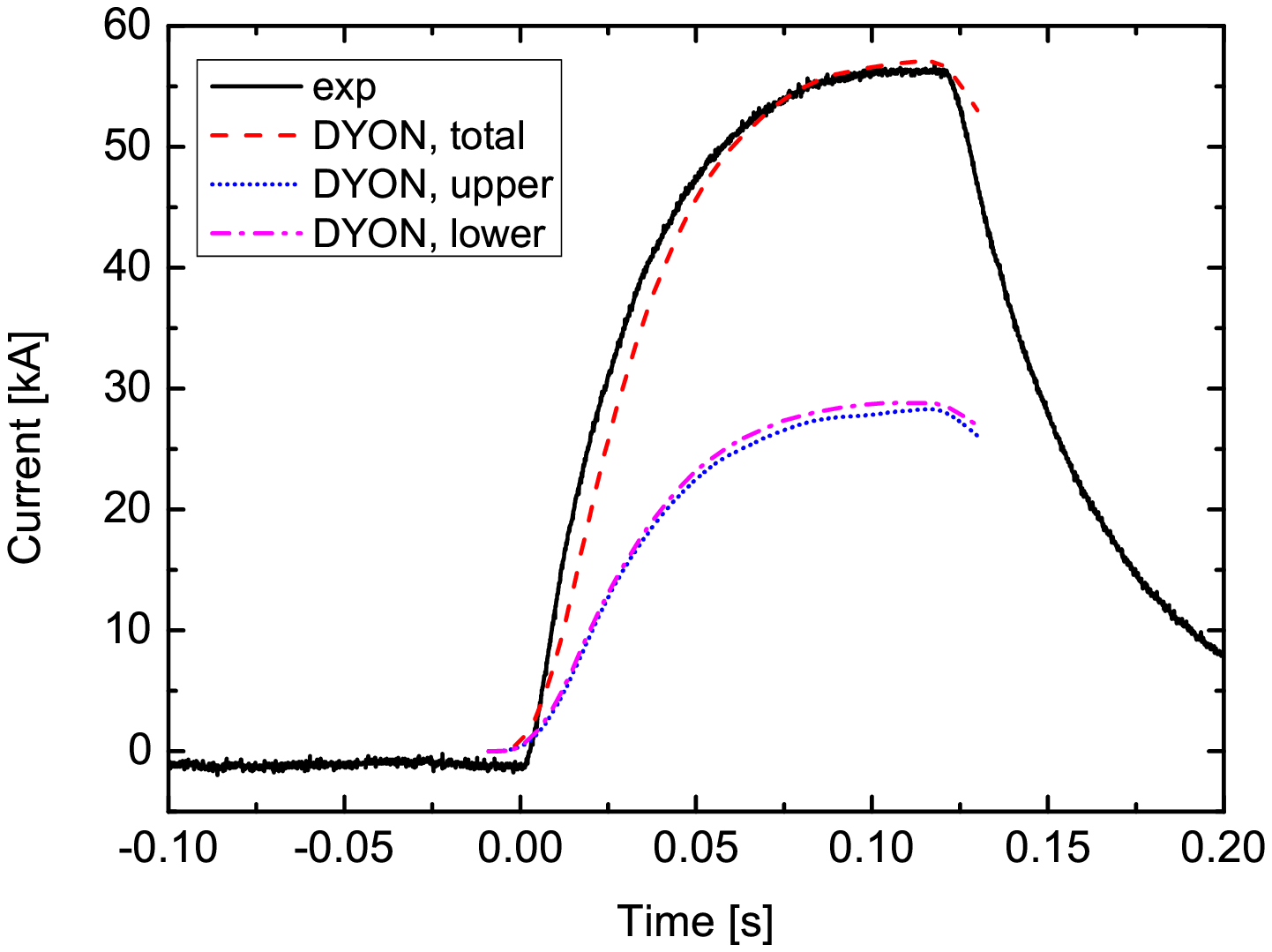}
    \caption{Comparison between total toroidal eddy current measured by rogoski coil in experiment (black solid) and that flowing through divertors in DYON prediction (red dashed). Blue and magenta curves means eddy current flowing through upper ans lower divertors, respectively.}
    \label{fig:vacuum_rc03}
\end{figure}

Passive materials are stainless steel and so their nominal resistivity is $7.76 \cdot 10^{-7} \ \Omega \cdot m$. For the effective resistivity, we multiply it by $0.4$ for divertors and $0.2$ for some passive structures colored red (labelled as isUL, osUL, isLL and osLL), respectively. We choose a vacuum discharge ($\#$30712) in KSTAR as the vacuum optimization target. The multiplication factors used successfully reproduce the measured loop voltage at the five channels as demonstrated in Fig. \ref{fig:vacuum_lv}. The five channels broadly covers a region inside the boundary, which indicates that the stray field calculation has also a good accuracy. Indeed, it's identified in Figs. \ref{fig:vacuum_null} that $B_z$ calculated by DYON-EM aligns well with FIST99 prediction. Note that DYON reliably predicts the total toroidal eddy current although it only considers the divertor toroidal currents as clarified in Fig. \ref{fig:vacuum_rc03}.

\subsection{DYON-RE comparison \label{ssec:DYON-RE}}
\begin{figure}[ht]
\centering
\includegraphics[width=0.5\textwidth]{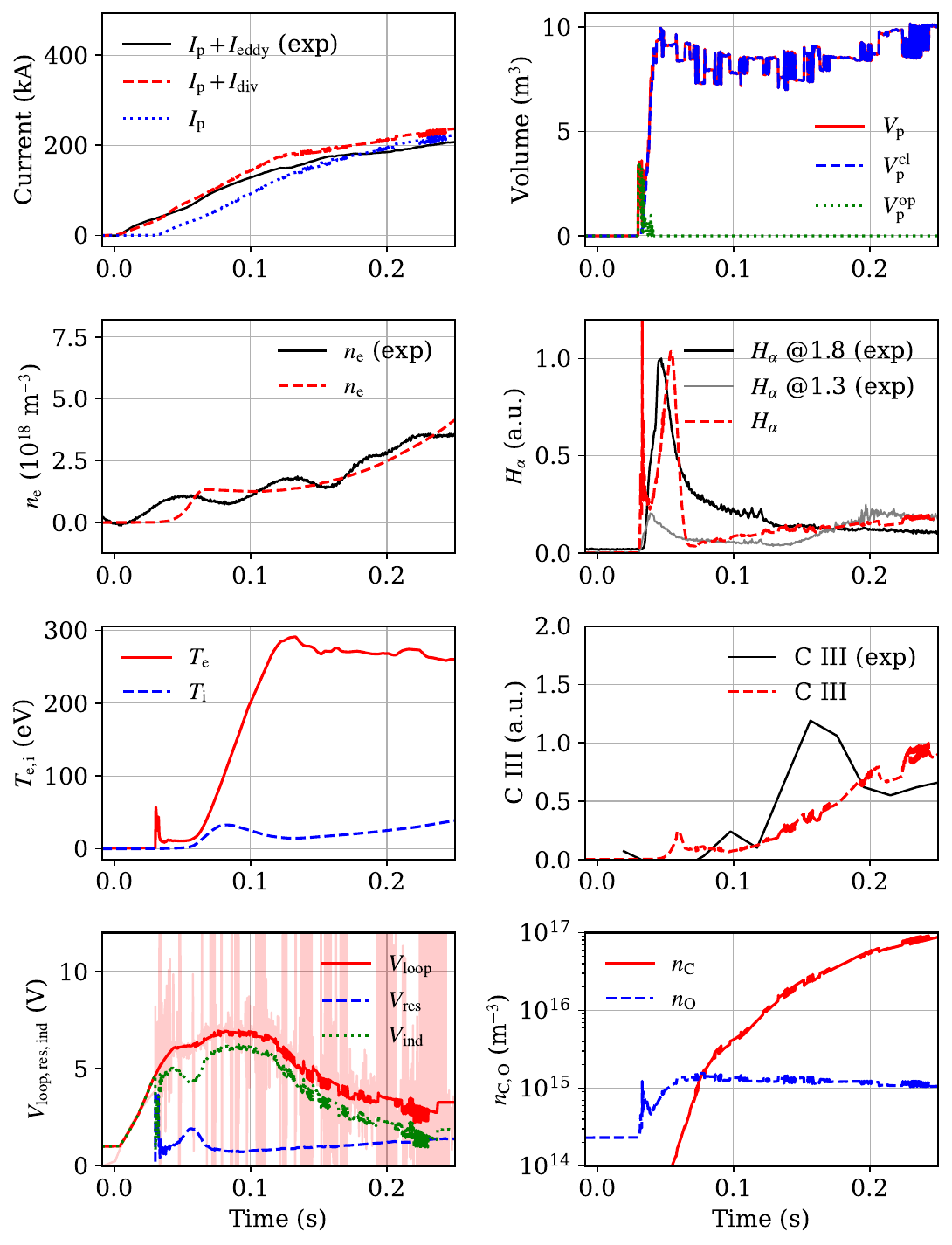}
\caption{Time evolution of plasma parameters for $\#$26031 (RE rich). 
Lines with a label "(exp)" in legend are the experimentally measured data and others are DYON-RE prediction. In the $H_\alpha$ panel, $@1.8$ and $@1.3$ denote the radial positions corresponding to the core ($1.8 \, \mathrm{m}$) and the inboard center ($1.3 \, \mathrm{m}$), respectively.}
\label{fig:plz_26031}
\end{figure}

\begin{figure}[ht]
\centering
\includegraphics[width=0.5\textwidth]{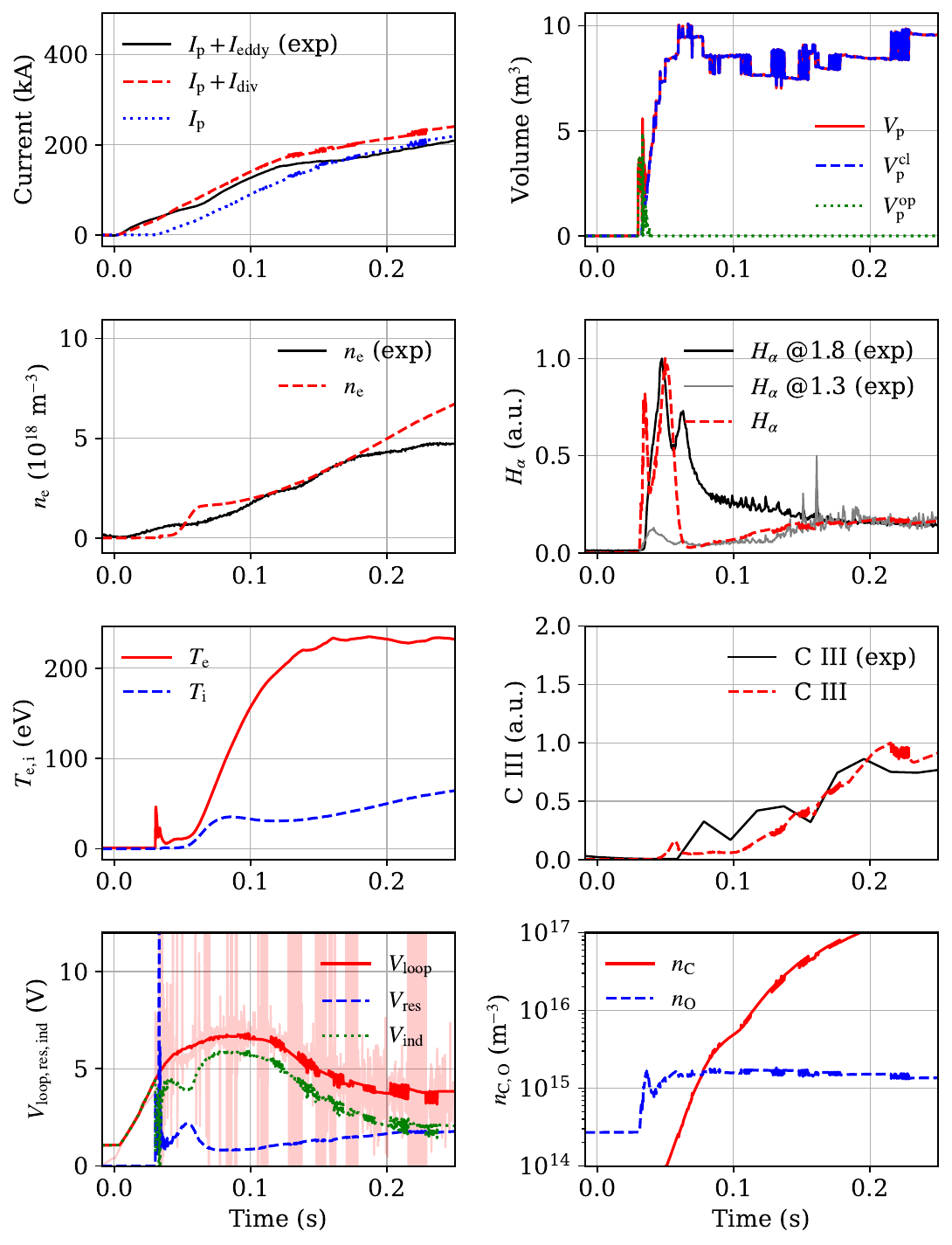}
\caption{Same to Fig. \ref{fig:plz_26031} for $\#$27340 (RE scarce).}
\label{fig:plz_27340}
\end{figure}

\begin{figure}[ht]
\centering
\includegraphics[width=0.5\textwidth]{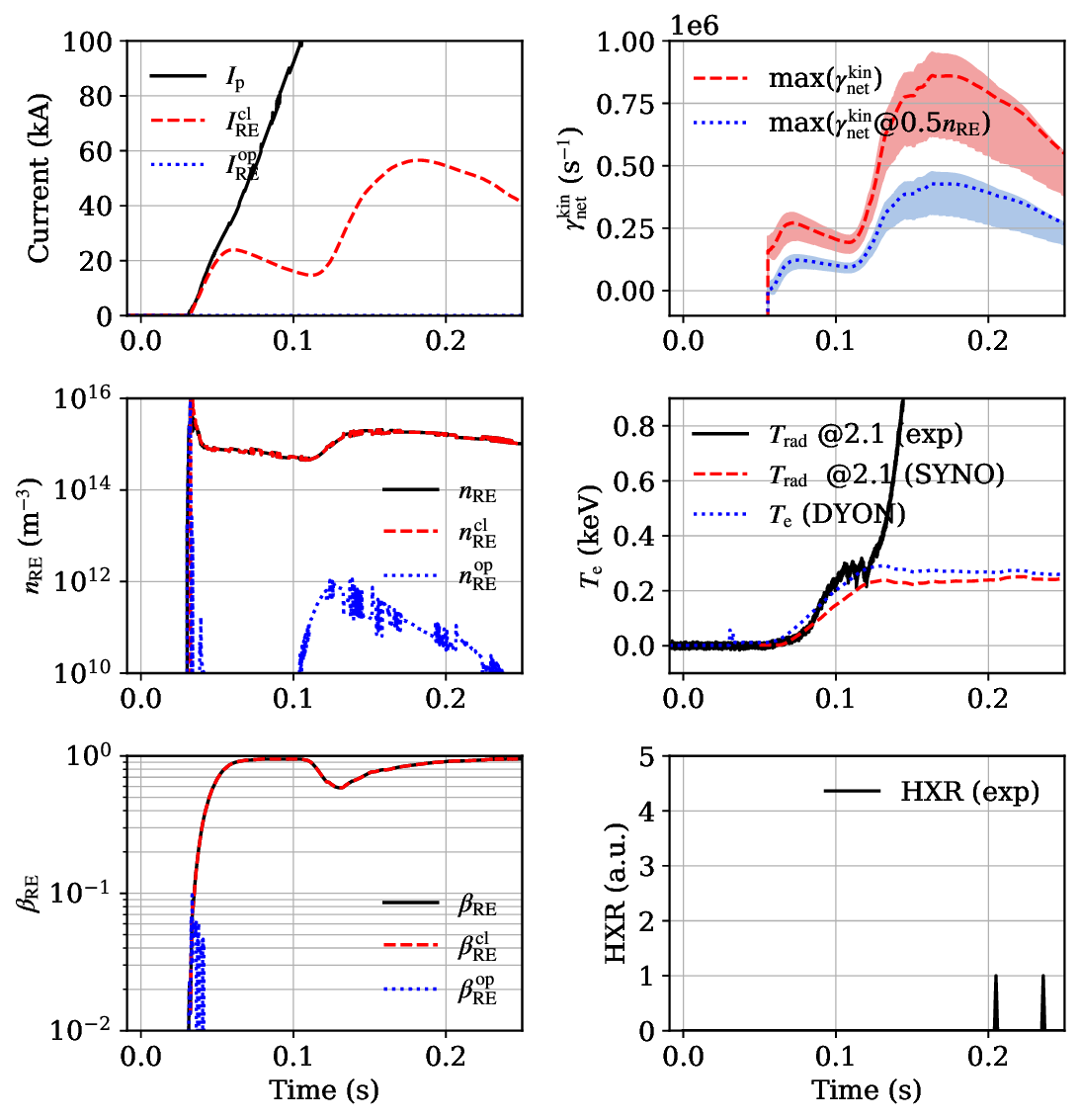}
\caption{Time evolution of runaway parameters for $\#$26031 (RE rich). 
(exp) in legend denotes the experimentally measured data and (SYON) is the SYNO simulation result. Right top panel shows the KIAT simulation results with $\gamma_{RE}=4$, where shaded area show a scan within $\gamma_{RE} \in [1.2, 5.5]$ and blue curve is obtained by using a half of $n_{RE}$. Others are DYON-RE prediction. In the $T_e$ panel, $@2.1$ denotes the radial position ($2.1 \, \mathrm{m}$).}
\label{fig:re_26031}
\end{figure}

\begin{figure}[ht]
\centering
\includegraphics[width=0.5\textwidth]{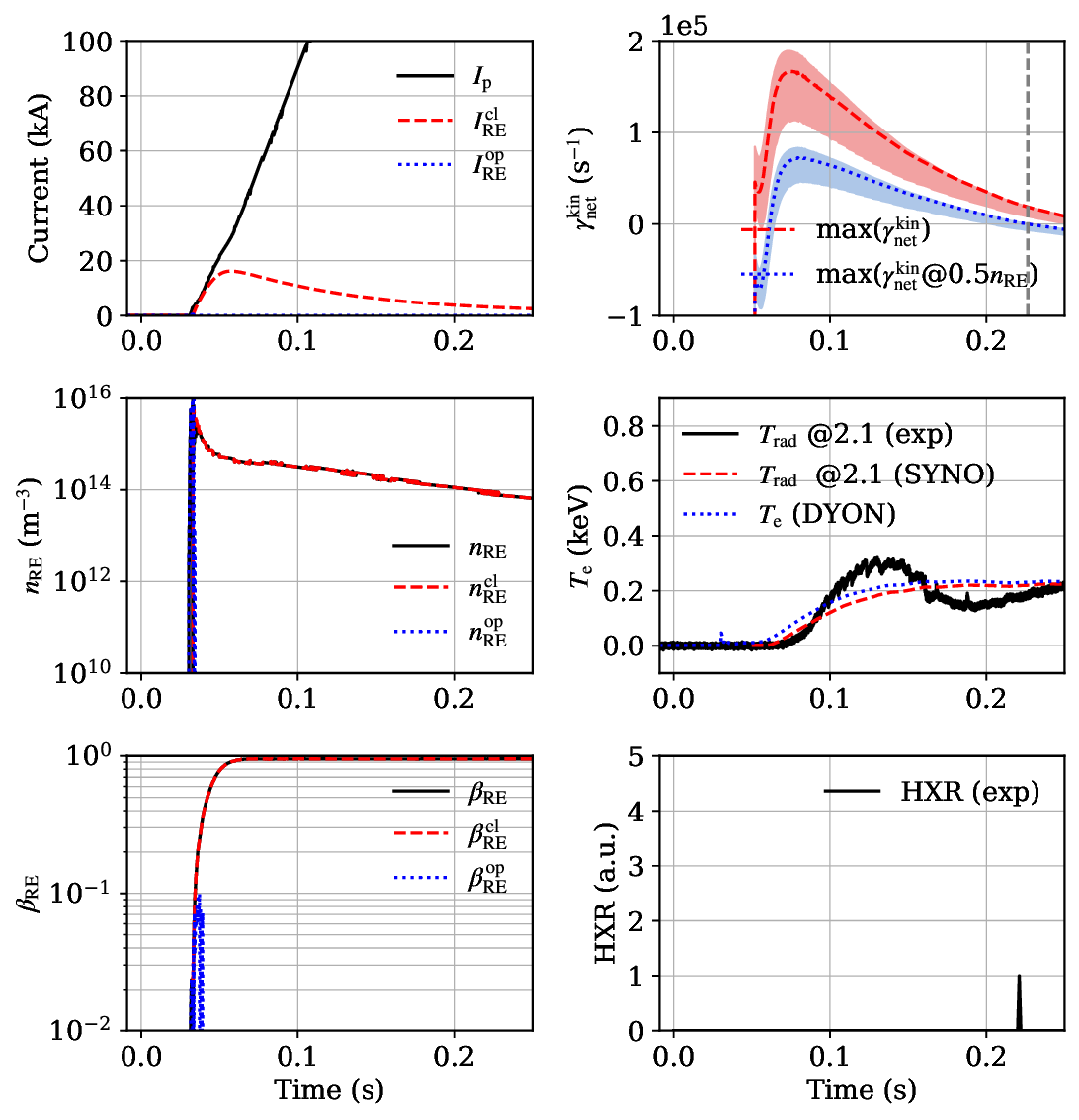}
\caption{Same to \ref{fig:re_26031} for $\#$27340 (RE scarce).}
\label{fig:re_27340}
\end{figure}

In this work, we do not attempt to determine the free parameters ($c_T$, $c_{p_0}$, $c_{n\Lambda}$ and $c_{RR}|\frac{\tilde{b}_r}{B_\phi}|^{-2}$) through an independent methodology such as the BREAK simulation, nor to characterize how these parameter choices affect the discharge. Instead, we focus on whether the proposed model can reproduce the experimental observations. To address this, choices of free parameters are $c_T=2.5$ for atomic collisions with main species, $c_{p_0}=0.4$ and $c_{n\Lambda}=0.88$ as one of the possible sets of values that explains the plasma parameter evolutions and sign of startup runaway electrons in KSTAR. A systematic determination of $c_T$ from BREAK, together with an assessment of how the resulting choice affects the predictive accuracy of the plasma parameters, is left to a forthcoming DYON-EM validation study in KSTAR. Numerical scheme to preserve the runaway particle and momentum conservations is introduced in \ref{App1}.

Figures \ref{fig:plz_26031}-\ref{fig:re_27340} demonstrate the qualitative comparison results of DYON-RE. Evolutions of plasma parameters are shown in Figs \ref{fig:plz_26031},  \ref{fig:plz_27340} whereas those of runaway parameters are in Figs. \ref{fig:re_26031}, \ref{fig:re_27340}. $\#$26031 and $\#$27340 are selected to represent RE rich shot (in Figs. \ref{fig:plz_26031}, \ref{fig:re_26031}) and RE scarce shot (in Figs. \ref{fig:plz_27340}, \ref{fig:re_27340}), respectively. 

In simulations, we assume the same position control scenario from outboard limited ($0.1 \ s$) to inboard limited ($0.2 \ s$) and thereby set the same time evolution of carbon sputtering coefficient \cite{Lee2023PS}. Similarly, the effective prefill gas pressure is accounted for by the same $c_{p_0}=0.4$. However, the main-species recycling coefficient $Y_D^D$ is differently varied to follow the measured density evolution. The recycling coefficient was adjusted to mimic the measured electron density behavior by compensating the use of the lower prefill gas pressure and the rapid ionizing avalanche in simulation (recall discussion in Sec. \ref{sssec:p0}). We take its waveform from Eq. (2.17) of Ref. \cite{Kim2012NF}
\begin{equation}
    Y_D^D(t) = c_1 - c_2 \Big( 1 - \exp\Big( - \frac{t}{c_3} \Big) \Big),
\end{equation}
where the coefficients $c_1=0.8$, $c_2 = -0.4$, $c_3=0.1$ are found for 26031 and $c_1=1.3$, $c_2=0.25$, $c_3=0.1$ are for 27340. We evolve toroidal eddy current from $-0.09 \ s$ but consider the plasma behavior starting at the later time ($0.03 \ s$) for the following reason: in KSTAR, although the loop voltage is applied from $0 \ s$, the electron density grows slowly due to the self-generated electric field effects driving profile inhomogeneity \cite{Yoo2018Nat} (recall discussion in Sec. \ref{sssec:3.1.1}), resulting in a rise in the D$_\alpha$ signal at approximately t=0.03s. We therefore define this point as the starting point since DYON assumes the sufficient ionization and adopts an initial ionization fraction of $0.2 \%$ \cite{Kim2012NF}. For oxygen, we consider the initial density fraction $n_{O,0} / n_{D,0} \approx 0.1 \ \%$. But, for carbon, we only consider their influx from the wall so use $n_{C,0} \approx 0$.

Figures \ref{fig:plz_26031} and \ref{fig:plz_27340} show that the self-consistent DYON-RE well reproduces evolutions of the plasma current, electron density, D$_\alpha$ lines, C $III$ line. The timing of D$_\alpha$ peaks are initially close with the measurement at R=$1.8 \ m$. As plasmas become limited to the inboard wall, rise in DYON successfully captures that in the measurement at R=$1.3 \ m$. When $t\leq 0.25 \, s$ in KSTAR, a reliable magnetic equilibrium is unavailable. To compare the electron density, it is necessary either to convert the line-integrated density measured by the interferometer into a line-averaged form, or to express the simulated density in an equivalently averaged form. In n$_e$ comparison, we adopt the latter approach and redefine the electron density by dividing the total number of electrons to the KSTAR standard volume with the major radius R$_0 = \ 1.8 \ m$, and minor radius a=$0.5 \ m$.

In either RE rich ($\#$26031) or RE scarce ($\#$27340) discharges, DYON-RE predicts a significant fraction of initial runaway current attributed to the confinement of runaway seeds inside a local closed flux surface, as demonstrated in Figs. \ref{fig:re_26031} and \ref{fig:re_27340}. Indeed, the maximum of $I_{RE}$ is about $24 \ kA$ for $\#$26031 and  $16 \ kA$ for $\#$27340, respectively. Meanwhile, only in $\#$26031, does carbon impurity influx drive the strong primary generation during early rampup \cite{Lee2023NF} due to its lower density. Note that our model self-consistently gives rise to the mildly relativistic correction in the runaway current so the characteristic runaway velocity is far less than 1, i.e. $\beta \ll 1$ whenever the seed production is strong enough.

The ECE signal retains the runaway signatures. For the purpose of the qualitative comparison, in Figs. \ref{fig:re_26031}, \ref{fig:re_27340}, we use KIAT and SYNO \cite{Lee2026arXiv2}. As shown in Fig. \ref{fig:param_evol}, $E/E_c \gg 1$ and $Z_{eff}\sim O(1)$ are met in KSTAR startup, where the mildly relativistic runaway electrons would have an anisotropic pitch-angle distribution without wave particle interactions \cite{Connor1975NF}. The RE momentum distribution function is simplified as an exponential momentum spectrum with an average momentum $p_0$ and a Gaussian pitch-angle distribution with an angle spread $\theta_0$ \cite{Rosenbluth1997NF, Aleynikov2015NF}. This simplification cannot represent the enhanced pitch-angle distribution of REs due to resonant wave–particle interactions. However, it is sufficient to demonstrate that the onset of kinetic instability can be predicted at the linear level from the RE density calculated by DYON-RE, and that the measured ECE radiation temperature can be either comparable to or significantly exceed the reconstructed ECE radiation temperature, depending on the amplitude of the linear growth rate.

KIAT computes the maximum value of the net growth rate of $\gamma^{kin}_{net} \equiv \gamma_{drive}^{kin}-\gamma_{drive}^{damp}$, where $\gamma_{drive}^{kin}$ is the kinetic driving rate and $\gamma_{drive}^{damp}$ is the collisional damping rate \cite{Aleynikov2015NF}. We only consider the whistler wave and magnetized plasma wave resonating with a zero-Larmor radius particle with $\gamma_{RE}$ through the anomalous Doppler resonance in this analysis by attributing the anomalous increase in ECE to them. SYNO reconstructs the synthetic radiative temperature $T_{rad}$ under the standard definition of $n_e$ and $n_{RE}$ (the total number of particles per the KSTAR standard volume) and DYON's $T_e$ only accounting for the second extraordinary mode. Also, SYNO considers a toroidal magnetic field scaling as $1/R$ and the reflection coefficient as $0.95$ without mode conversion. We draw the KIAT result for $\theta_0=0.1$ and $\gamma_{RE}\equiv\sqrt{p_{RE}^2+1}=4$ as curves and the scan results in $\gamma_{RE} \in [1.2,5.5]$ as shaded area in right top panel. The blue curve represents max($0.5 \ \gamma^{kin}_{drive} - \gamma^{coll}_{damp}$). In the SYNO analysis, the maximum $T_{rad}$ within the scan range of $\gamma_{RE} \in [1.2,5.5]$ is shown as red curve in right middle panel. In RE rich discharge ($\#$26031), the measured $T_{rad}$ in experiment initially follows the SYNO prediction but soars rapidly after $0.12 \ s$ reaching beyond $12 \ keV$, suggesting that RE parallel kinetic energy is converted to their perpendicular energy due to strong pitch-angle scattering enhanced by nonlinear kinetic instability and thereby $\gamma^{kin}_{drive} \gg \gamma^{coll}_{damp}$ during the early rampup. In fact, the $\gamma^{kin}_{net}$ panel in Fig. \ref{fig:re_26031} clarifies that the maximum $\gamma^{kin}_{net}$ remains above $0.1 MHz$, far higher than $\#$27340.

By contrast, in RE scarce discharge ($\#$27340, see Fig. \ref{fig:re_27340}), the measured $T_{rad}$ agrees well with the SYNO prediction in magnitude but shows sawtooth-like bursting behaviors, suggesting that $\gamma^{kin}_{drive}$ is comparable to $\gamma^{coll}_{damp}$ but slightly higher. These behaviors disappear at $0.2 \ s$, from which the kinetic instability is forbidden. Indeed, the blue curve (=max($0.5 \ \gamma^{kin}_{drive} - \gamma^{coll}_{damp}$)) shows that its maximum is positive but below $0.1 MHz$, and subsequently becomes negative, i.e. the kinetic stabilization. While the analysis performed by KIAT is 'local', but in a reality waves propagate in plasmas so the 'global' convective instability onset needs either higher runaway density or higher electron temperature \cite{Aleynikov2015NF}. Therefore, although the red curve predicts $\gamma^{kin}_{net}>0$ in Fig. \ref{fig:re_27340}, the result obtained by assuming a lower RE density (blue curve) shows better agreement with the experiment, which can be considered more reasonable from a qualitative perspective.

\subsection{Origin of RE content classification and role of the seed RE across CFSF\label{ssec:noEM}}
\begin{figure}[ht]
\centering
\includegraphics[width=0.5\textwidth]{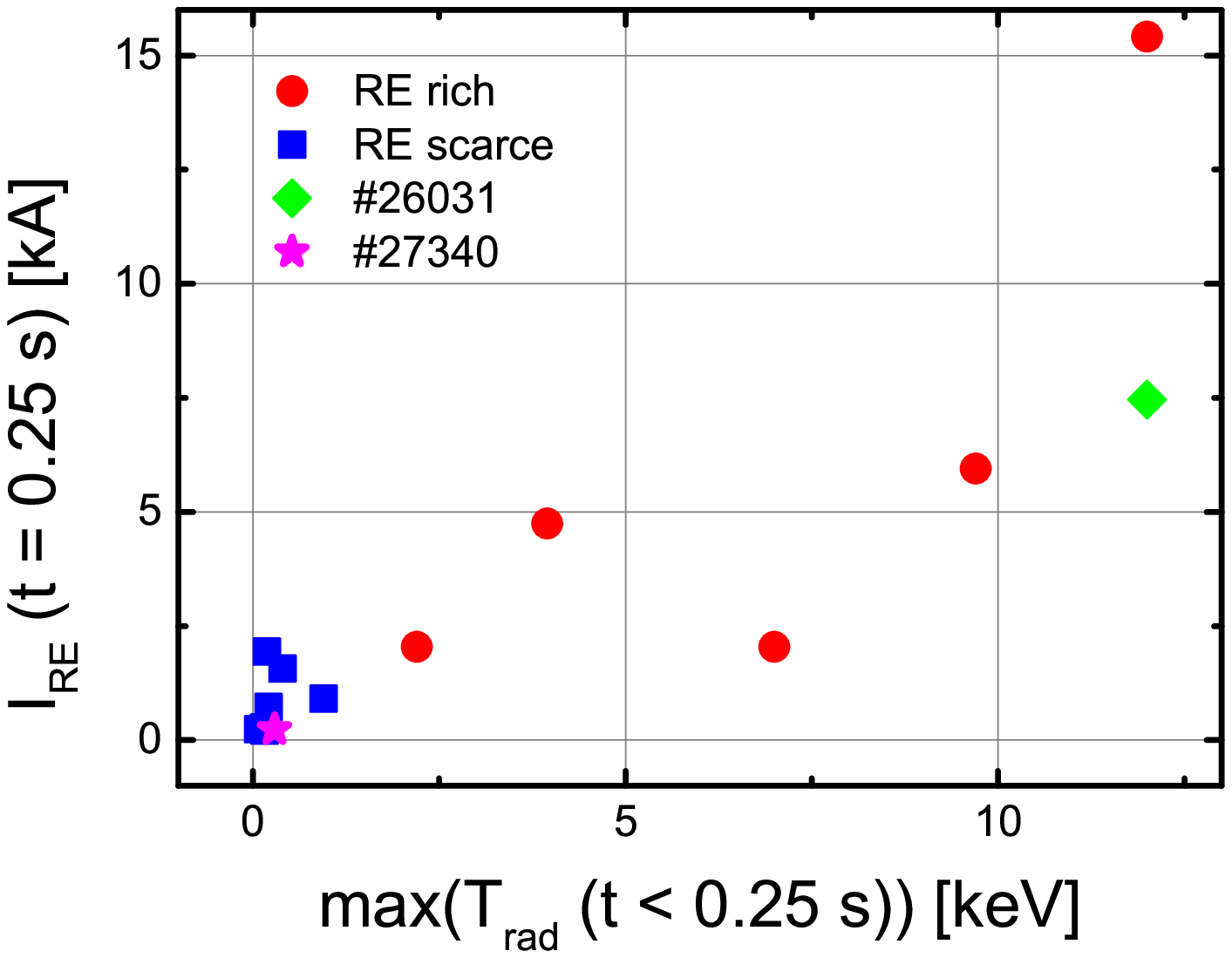}
\caption{Scatter plot of the maximum runaway current density $I_{RE}$ at $t = 0.25 \, \mathrm{s}$ against the maximum radiative temperature $\max(T_{rad})$ for $t < 0.25 \, \mathrm{s}$, for the RE-rich (red circle) and RE-scarce (blue square) discharges. Here, DYON-RE uses the macroscopic runaway transport model without the electromagnetic model. Green diamond and magenta star denote $\#$26031 (RE rich) and $\#$27340 (RE scarce), respectively.}
\label{fig:noEM}
\end{figure}

In Sec. \ref{ssec:DYON-RE}, the comparison was carried out for only two representative discharges, showing that whether strong runaway formation occurs during the early current rampup affects the difference between RE rich and RE scarce shots. To examine its generality across multiple discharges, we performed DYON-RE simulations for the discharges shown in Fig. \ref{fig:Bif}. As suggested in Sec. \ref{ssec:LCFSF}, the macroscopic RE transport model would be sufficient if the early current rampup were the phase that governed the separation between the two groups. We therefore use the macroscopic description for RE confinement without the electromagnetic model, where we use the interpolation (Eq. \ref{eq:tau_RE_gloabal}) with $I_{RE,ref}=I_{ref}/15$ and assume a good perpendicular confinement $\tau_{RE,\perp} = 2 \, \mathrm{s}$ (see Sec. 5.4 of Ref. \cite{Lee2025phD} for the detail).

Figure \ref{fig:noEM} demonstrates a positive correlation between the experimental radiative temperature and simulated runaway current although these two quantities do not directly correspond to each other. This qualitatively supports that the origin of the RE content classification is the runaway electron generation during the early current rampup phase.

On the other hand, as expected, describing the startup runaway generation across the CFSF phase is found to require the dual population description coupled with the CFSF model. Indeed, for $\#$27340, the final runaway current is only $300 \, \mathrm{A}$, far less than the prediction of $\sim 2.5 \, \mathrm{kA}$ obtained with the coupled dual population treatment. In the no-EM run, no region of positive growth rate $\gamma_{net}^{kin}>0$ exists, and the mildly bursting behavior in ECE could not be explained.

\section{Discussion and conclusions \label{sec:discussion}}
Startup RE current has been often modelled by assuming $v_{RE}=c$ \cite{Gribov2018EPS, Hoppe2022JPP, Matsuyama2022NF, deVries2025NF}. In this work, we developed the model of mildly relativistic REs to describe startup REs and verified it by showing agreement with the kinetic simulation. This model can be simplified to the reduced-kinetic form while reasonably preserving the mildly relativistic feature ($v_{RE}\neq c$). In DYON-RE, we implement the reduced model into DYON \cite{Kim2022NF}. The binary nature of collisions between fast electrons and neutral \cite{Lee2024PRL, Lee2026JPP} is also considered in the primary RE generation as an ad-hoc manner. Furthermore, two RE populations are introduced in an open and closed magnetic configuration and their confinements are separately described to account for the transition in dominant confinement mechanisms during CFSF. In conclusion, our treatment primarily improves startup RE modellings by elaborating the fluid RE current and resolving the startup RE transport.

This work leads to two main advances building upon previous efforts taking the first steps towards validation in Ref \cite{deVries2025NF}. First, the model itself has been improved: DYON-RE now incorporates the full-electromagnetic model and mildly relativistic RE model, enabling it to capture the listed features in the above paragraph. Second, a qualitative comparison with the measured ECE emission characteristics has been performed with additional dedicated analyses using the KIAT and SYNO codes. The linear analysis of kinetic instability and synthetic reconstruction of the radiation temperature indirectly show that the evolutions of RE density predicted by DYON-RE are consistent with the measured ECE signatures in KSTAR ohmic discharges: It qualitatively explains strong destabilization, mildly bursting behavior and stabilization of modes. 
Although the obtained solution cannot be regarded as unique, we demonstrate that a solution exists which can qualitatively explain the experimental observations. However, since this is not a \textit{quantitative} validation in a strict sense, particular caution is required when performing predictive simulations for future devices such as ITER or CPD.

The classification of ohmic startups by their runaway content is not a phenomenon observed only in KSTAR. A recent multi-machine study confirmed that this trend is observed generically across various devices\cite{Vries2023NF}, and the electron density evolution has been suggested as important for understanding it \cite{Vries2020PPCF, deVries2025NF}. In this context, our multi-discharge simulations support the interpretation that such RE content classification is a consequence of the runaway generation during the early current rampup phase with low electron density.

An unintended takeaway of this work is the importance of modelling the plasma–wall interaction that includes the effective compensation of the lowered prefill gas pressure, supporting the same conclusion discussed in Sec. 3.1.5 of Ref. \cite{deVries2025NF}. As proposed in Ref. \cite{Devries2020PPCF}, a density control during the early rampup phase has been suggested as a possible strategy for mitigating startup REs, and Ref. \cite{Hoppe2022JPP} showed that this approach can provide meaningful mitigation if properly implemented. However, trends observed in KSTAR ohmic discharges indicate that, even with similar prefill windows, the density evolution can vary significantly depending on the wall condition in superconducting tokamaks such as ITER. This implies that, for density control via gas puffing to be effective, the wall-induced particle source must be understood and quantified. An amount of out-gassing from the wall itself might be reduced when the carbon wall used in the analyzed KSTAR discharges is replaced with the metal wall \cite{Devries2013PPCF} but how to compensate the lowered prefill gas pressure is still unclear.

In this study, we only suggest that the wall condition plays an important role in startup RE formation in KSTAR, but we do not provide a predictive framework for incorporating this effect. One possible approach is to scan a constant recycling coefficient, as done in Ref. \cite{Matsuyama2022NF}. However, our observations indicate that reproducing the density evolution may require accounting for its temporal variation, which introduces additional uncertainty; the constant recycling coefficient may be possible for a certain set of ($c_{p_0}$ and $c_T$). Therefore, the other conclusion of this work is that, without properly accounting for the wall response, it is difficult to reliably describe runaway electron dynamics during the early rampup phase.

\section*{Acknowledgement}
Y. Lee is grateful to Dr. Min-Gu Yoo for fruitful discussion. This research was supported by R\&D Program of "Optimal Basic Design of DEMO Fusion Reactor, CN2602-4" through the Korea Institute of Fusion Energy (KFE) funded by the Government funds.

\appendix
\section{Numerical scheme for the conservation laws \label{App1}}
The full electromagnetic plasma burn-through model has been developed to consider the particle and energy conservations \cite{Kim2022NF}. Equation (8) of Ref. \cite{Kim2022NF} has a typo that missed the change in electron and ion energy density attributed to the evolving plasma volume, which we considered in our simulation by subtracting $\frac{3}{2}n_e T_e \frac{d}{dt} \ln V_p$ and $\frac{3}{2}n_i T_i \frac{d}{dt} \ln V_p$ in the right hand sides, respectively.

In numerical implementation, the plasma volume is evaluated at coarse time intervals, rather than at every time step, to reduce the computational cost of DYON-EM. As a result, the temporal evolution of $V_p$ and $dV_p / dt$ between successive Townsend evaluations is not fully resolved, complicating the preservation of conservation laws. To address this issue, we explicitly and self-consistently update $V_p$ and $dV_p/dt$ in a smooth manner, but this introduces a one-time-step temporal lag.

Suppose that $V_p = V_0$ is assessed at $t=t_0$ and $V_p = V_1$ at $t=t_1$. The method we propose is to smoothly extrapolate its transition. Let $V_p^+ = (V_0 + V_1)/2$, $V_p^- = (V_1-V_0)/2$, $t^+ = (t_1+t_2)/2$ and $t^- = (t_2-t_1)/2$, where $t_2 -t_1 = t_1-t_0$. Extrapolating $V_p$ from $t=t_1$ to $t=t_2$ with vanishing gradient condition at the boundaries yields,
\begin{equation}
    V_p (t) = 
        - \frac{V_p^-}{2(t^-)^3} \Big( (t-t^+)^3 - 3 (t^-)^2(t-t^+) \Big) + V_p^+ \label{eq:Vp_smoo}
\end{equation}
and its derivative reads
\begin{equation}
    \frac{dV_p}{dt} (t) = 
    - \frac{3 V_p^-}{2(t^-)^3} \Big( (t-t^+)^2 - (t^-)^2\Big) + V_p^+.  \label{eq:dVp_smoo}
\end{equation}

Although this method resolved the conservation issues, the evolution of plasma volume is not synchronized with its actual evaluation. Hence, the time interval used to estimate $V_p$ should be carefully chosen: a long interval leads to unreliable $V_p$-evolution, whereas a short interval may induce nonphysical numerical oscillations. In other words, a prior assessment of an appropriate time interval is required when employing this scheme. In this work, we found the proper interval as $0.3 \ ms$ through a trial-and-error process.

\section*{References}

\bibliographystyle{unsrt}
\bibliography{ref}

\end{document}